\documentclass{article} 
\usepackage{iclr2027_conference,times}

\usepackage{amsmath,amsfonts,bm}

\def\eqref#1{equation~\ref{#1}}

\def\1{\bm{1}}

\DeclareMathAlphabet{\mathsfit}{\encodingdefault}{\sfdefault}{m}{sl}
\SetMathAlphabet{\mathsfit}{bold}{\encodingdefault}{\sfdefault}{bx}{n}

\usepackage{hyperref}
\usepackage{url}

\usepackage{hyperref}

\usepackage{xspace}
\usepackage{xstring}

\usepackage{booktabs}
\usepackage{multirow}
\usepackage{tabularx}
\usepackage{threeparttable}

\usepackage[utf8]{inputenc}
\usepackage{xcolor}
\usepackage{listings}
\usepackage{caption}
\usepackage{varwidth}

\usepackage[most]{tcolorbox}

\usepackage[T1]{fontenc}

\usepackage{placeins}
\usepackage{float}  

\usepackage{booktabs}
\usepackage{tabularx}
\usepackage{listings}
\usepackage{xcolor}
\usepackage{tcolorbox}
\usepackage{caption}
\usepackage{subcaption}
\usepackage{amsmath}

\newcommand{\sys}{\mbox{\textsc{ICall-Bench}}\xspace}

\newcommand{\PP}[1]{\noindent{\bf \IfEndWith{#1}{.}{#1}{#1.}}}
\newcommand{\IPP}[1]{\smallskip\noindent{\textit{\IfEndWith{#1}{.}{#1}{#1.}}}}

\definecolor{keywordgreen}{rgb}{0.13, 0.55, 0.13} 
\definecolor{typecolor}{rgb}{0.65, 0.16, 0.16}    
\definecolor{commentblue}{rgb}{0.0, 0.5, 0.5}   
\definecolor{numbergray}{rgb}{0.5, 0.5, 0.5}

\lstdefinestyle{promptstyle}{
    basicstyle=\ttfamily\footnotesize,
    breaklines=true,
    breakatwhitespace=false,
    columns=fullflexible,
    keepspaces=true,
    showstringspaces=false,
    aboveskip=0pt,
    belowskip=0pt,
    lineskip=-0.5pt,
    frame=none
}

\newtcolorbox{promptbox}{
    colback=white,
    colframe=black,
    boxrule=0.8pt,
    arc=2mm,
    outer arc=2mm,
    left=3mm,
    right=3mm,
    top=2mm,
    bottom=2mm,
    boxsep=0pt
}

\newcommand{\prog}[1]{\textcolor{blue}{#1}}

\title{\sys: Benchmarking LLM Interprocedural Reasoning in Binary Programs}

\author{
\begin{tabular}{@{}c@{\hspace{0.02\textwidth}}c@{}}
\begin{minipage}[t]{0.48\textwidth}
\vspace{0pt}
\centering
\normalfont
\textbf{Shenglong Yao}\\
Georgia Institute of Technology\\
\texttt{syao84@gatech.edu}
\end{minipage}
&
\begin{minipage}[t]{0.48\textwidth}
\vspace{0pt}
\centering
\normalfont
\textbf{Zirui Liu}\\
University of Massachusetts Amherst\\
\texttt{zliu@umass.edu}
\end{minipage}
\\[1em]
\begin{minipage}[t]{0.48\textwidth}
\vspace{0pt}
\centering
\normalfont
\textbf{Xiangchi Yuan}\\
Georgia Institute of Technology\\
\texttt{xyuan300@gatech.edu}
\end{minipage}
&
\begin{minipage}[t]{0.48\textwidth}
\vspace{0pt}
\centering
\normalfont
\textbf{Wenke Lee}\\
Georgia Institute of Technology\\
\texttt{wenke@cc.gatech.edu}
\end{minipage}
\end{tabular}
}

\iclrfinalcopy

\begin{document}

\maketitle
\lhead{} 

\begin{abstract}

Recent benchmarks show that large language models (LLMs) achieve strong performance in both function-level binary understanding and end-to-end binary reverse engineering tasks, but whether LLMs can reason over interprocedural semantics in compiled binaries remains unclear.
Function-level benchmarks offer limited visibility into cross-function relationships, while whole-program benchmarks typically evaluate final outcomes of tasks that may be solvable from local clues alone, making it difficult to disentangle interprocedural reasoning from exploration, tool use, execution, and retrieval.
To address this limitation, we introduce \sys, a benchmark that directly evaluates whether LLMs and agents can understand and reason over interprocedural semantics in binary programs, using indirect-call target recovery as a measurable probe.
\sys covers 50 complete real-world C/C++ programs spanning different complexity levels and multiple optimization configurations, with an average of 74,746 lines of code and 941 functions per program, enabling analysis of reasoning at whole-program scale and under diverse binary representations. 
We further introduce three analysis regimes spanning model-only reasoning, static tool-augmented agents, and Internet-enabled open-world agents to examine how intrinsic reasoning, active information acquisition, and external knowledge each contribute to interprocedural binary analysis.
We evaluate five recent LLMs and their corresponding agents.
Our results reveal a wide separation across current models and agents, with program-level verifiable success rates ranging from 23\% to 81\%.
They also show that greater information access does not consistently lead to better results, with particularly limited gains on the hardest cases, where deeper interprocedural reasoning remains the primary bottleneck.
Uneven performance across models and frequent verifiable failures even under an optimistic evaluation indicate that robust binary interprocedural reasoning remains an open challenge.

\end{abstract}

\section{Introduction}
\label{sec:intro}
LLMs and LLM-powered agents have recently shown growing capabilities in binary analysis and reverse engineering.
Existing benchmarks for LLM-based binary analysis fall into two categories. 
Function-level benchmarks \citep{gao2025decompilebench, shang2025binmetric, won2026rebench, NEURIPS2025_079cf13a} evaluate tasks such as decompilation, symbol recovery, and type recovery, while whole-program agentic reverse-engineering benchmarks \citep{agentrebench2026, david2026crackmebench, chen2026crebench, spence2026next} evaluate end-to-end outcomes such as artifact recovery, or challenge completion with tool-using agents. 
Although these benchmarks show progress in local binary understanding and autonomous reverse engineering, they provide limited visibility into a specific capability that lies between the two: interprocedural reasoning. 
Strong function-level performance does not reveal whether a model can connect information distributed across multiple functions.
Meanwhile, success in agentic whole-program tasks does not directly reflect interprocedural reasoning, as it is entangled with other capabilities, including tool use, execution, retrieval, and search, and can usually be achieved by focusing on a small set of task-relevant code regions rather than reconstructing a coherent view of the program’s cross-function structure.
This capability matters as many downstream binary-analysis tasks require interprocedural reasoning, including control-flow integrity (CFI), taint analysis and vulnerability discovery.
We therefore ask: can models capable of function-level binary understanding also reason about the interprocedural  semantics?

To study this question, we use binary indirect-call target recovery as a diagnostic probe of binary interprocedural reasoning. 
Unlike direct calls, indirect calls resolve their targets only at runtime, making precise static resolution a challenging and still-open problem requiring interprocedural data- and control-flow analysis, while the resulting callsite-to-target relations remain mechanically verifiable through execution.
Therefore, binary indirect-call recovery provides a compact and diagnostic probe of binary interprocedural reasoning where each prediction requires recovering a latent relation between a callsite and its possible callees.

Having selected indirect-call recovery as the probe, we design the benchmark around two questions.
\textbf{First, how robustly can current systems do interprocedural reasoning across diverse real-world program settings?}
Pre-selecting a few relevant functions risks reducing interprocedural analysis to local reasoning and renders it meaningless for large whole-program tasks. 
We instead evaluate complete real-world programs spanning different sizes, programming languages, and optimization levels, with large function graphs and numerous indirect-call sites. 
This diversity allows us to assess how robustly models recover cross-function semantics across realistic program settings.
\textbf{Second, how does interprocedural reasoning change as AI systems gain increasingly powerful mechanisms for acquiring information?}
We evaluate the same task under three analysis regimes: model-only reasoning, static tool-augmented agents, and Internet-enabled open-world agents.
These settings progressively allow systems to reason over provided evidence, actively inspect the binary, and incorporate external knowledge.
The Internet-enabled open-world setting reflects practical reverse engineering’s use of repositories, documentation, and related implementations, while also raising the risk of benchmark contamination. 
Indirect-call recovery is comparatively robust to such contamination because even access to corresponding source code rarely reveals the binary-level callsite-to-target relations directly.
Agents must still account for compilation and optimization and ground retrieved information in the specific binary.
This design therefore lets us evaluate not only intrinsic reasoning ability, but also how effectively agents combine binary analysis, active investigation, and external knowledge in realistic binary analysis workflows.

Based on this design, we construct \sys with 100 evaluation instances spanning C/C++ programs, diverse program sizes, and multiple compiler optimization levels. 
Each program contains, on average, 74,746 lines of code and 941 functions.
We apply \sys to five recent LLMs and their corresponding agents across three analysis regimes to recover indirect-call targets from the program instances.
Under the default static tool-augmented agent regime, performance varies widely across LLMs, with program-level verifiable success rates ranging from 23\% to 81\%, while compiler optimization has relatively little impact on performance.
Static tool augmentation enables agents to consistently pass more tasks than model-only reasoning, whereas Internet-enabled open-world access has a heterogeneous effect across models, ranging from a 10\% decrease to an 11\% improvement.
These results show that active analysis substantially strengthens interprocedural reasoning, but effectively exploiting additional information remains a key challenge.

In summary, we make the following key contributions:
\begin{itemize}
\item We formulate binary interprocedural reasoning as an explicit capability for LLM-based binary analysis, and introduce indirect-call target recovery as a concrete diagnostic probe.

\item We construct \sys with 100 real-world binaries and evaluate under three regimes spanning model-only, static tool-augmentation, and Internet-enabled open-world.

\item We evaluate five recent LLMs, revealing substantial capability gaps, strong gains from static tool augmentation and uneven effects of open-world access.

\end{itemize}

\section{Related Work}
\subsection{LLM Benchmarks for Binary Analysis and Reverse Engineering}

\begin{table*}[t]
\centering
\scriptsize
\setlength{\tabcolsep}{3.2pt}
\renewcommand{\arraystretch}{1.12}

\caption{
Comparison with Benchmarks for LLM-based Binary Analysis and
Reverse Engineering.
}
\label{tab:benchmark_comparison}

\resizebox{\textwidth}{!}{
\begin{tabular}{
    c
    c
    c
    c
    c
    c
    c
    c
    c
}
\toprule

\multirow{2}{*}{\textbf{Benchmark}} &
\multirow{2}{*}{\textbf{Task}} &
\multirow{2}{*}{\textbf{Level}} &
\multicolumn{3}{c}{\textbf{Scale \& Complexity}\textsuperscript{a}} &
\multicolumn{3}{c}{\textbf{Analysis Regime}} \\

\cmidrule(lr){4-6}
\cmidrule(lr){7-9}

&
&
&
\textbf{\# Prog.} &
\textbf{\# Eval. Units} &
\textbf{Avg. LOC/Prog.\textsuperscript{b}} &
\textbf{Model} &
\textbf{Tools} &
\textbf{Internet\textsuperscript{c}} \\

\midrule

DecompileBench~\citep{gao2025decompilebench}
&
Decompilation
&
Function
&
130
&
23.4K funcs.
&
--
&
$\checkmark$
&
$\times$
&
$\times$
\\

Decompile-Bench~\citep{NEURIPS2025_079cf13a}
&
Decompilation
&
Function
&
121
&
70K funcs.
&
--
&
$\checkmark$
&
$\times$
&
$\times$
\\

BinMetric~\citep{shang2025binmetric}
&
Decompilation
&
Function
&
20
&
1K funcs.
&
--
&
$\checkmark$
&
$\times$
&
$\times$
\\

REBench~\citep{won2026rebench}
&
Name \& type recovery
&
Function
&
96
&
40K funcs.
&
--
&
$\checkmark$
&
$\times$
&
$\times$
\\

\midrule

AgentRE-Bench~\citep{agentrebench2026}
&
General RE
&
Whole binary
&
13
&
13 bins.
&
63.5
&
$\times$
&
$\checkmark$
&
$\times$
\\

CrackMeBench~\citep{david2026crackmebench}
&
General RE
&
Whole binary
&
12
&
12 bins.
&
36.1
&
$\times$
&
$\checkmark$
&
$\times$
\\

CREBench~\citep{chen2026crebench}
&
Crypto RE
&
Whole binary
&
48
&
432 bins.
&
526.2
&
$\times$
&
$\checkmark$
&
--
\\

SRE-Bench~\citep{spence2026next}
&
General RE
&
Whole binary
&
19
&
262 bins.
&
16.9K
&
$\times$
&
$\checkmark$
&
$\times$
\\

\midrule

\textbf{\sys (ours)}
&
\textbf{Indirect-call recovery}
&
\textbf{Interprocedural}
&
\textbf{50}
&
\textbf{100 bins}
&
\textbf{74.7K}
&
$\checkmark$
&
$\checkmark$
&
$\checkmark$
\\

\bottomrule
\end{tabular}
}

\vspace{2pt}
\begin{minipage}{0.99\textwidth}
\footnotesize
\textsuperscript{a} All scale and complexity statistics use author-reported statistics for their instances in the actual evaluation.

\textsuperscript{b} Avg. LOC/Prog. is omitted for function-level benchmarks, where program-level size does not directly characterize the complexity of individual evaluation units.

\textsuperscript{c} "--" indicates that we found no explicit evidence showing whether Internet access was enabled or disabled.
\end{minipage}

\end{table*}

\autoref{tab:benchmark_comparison} summarizes two main categories of existing LLM binary-analysis benchmarks.
Function-level benchmarks, such as DecompileBench~\citep{gao2025decompilebench}, Decompile-Bench~\citep{NEURIPS2025_079cf13a}, BinMetric~\citep{shang2025binmetric} and REBench~\citep{won2026rebench}, evaluate tasks such as decompilation, type and name recovery, and other properties of individual functions.
These benchmarks assess local binary understanding but not cross-function relationship comprehension.
In contrast, recent agentic whole-program benchmarks, including AgentRE-Bench~\citep{agentrebench2026}, CrackMeBench~\citep{david2026crackmebench}, CREBench~\citep{chen2026crebench}, and SRE-Bench~\citep{spence2026next}, evaluate systems on end-to-end reverse-engineering tasks.
While this setting captures broader program-analysis capabilities, their final metrics conflate interprocedural reasoning with exploration, tool use, execution, search, and other capabilities. 
Moreover, solving these tasks often only requires identifying a relatively small set of task-relevant code regions rather than reconstructing a coherent view of the program's overall cross-function structure.
\sys occupies a complementary point between these categories: it preserves the complete-program context while providing a probe of interprocedural semantics understanding and reasoning.
Existing benchmarks also differ in scale, complexity, and external resources available to evaluated systems. 
Their datasets range from tens to more than one hundred ground-truth projects, with each instance spanning from tens to thousands of lines of code. 
As shown in \autoref{tab:benchmark_comparison}, \sys matches existing whole-program benchmarks in instance count while featuring more complex programs with higher average lines of code per instance. 
Analysis regimes for external resources vary as well: function-level benchmarks are typically restricted to model-only queries, whereas whole-program benchmarks rely on benchmark-specific agents with limited analysis tools and usually no Internet access.
In contrast, we evaluate the same task across model-only, static tool-augmented, and Internet-enabled open-world regimes, measuring the impact of active information acquisition and external knowledge.

\subsection{Call-Relation Reasoning Benchmarks}
Recent work has begun to evaluate call relations directly as a fine-grained probe of program reasoning.
TraceEval~\citep{li2026execution} assesses whether LLMs can recover execution-covered caller-callee relations from source code, using runtime call graphs as mechanically verifiable ground truth.
CG-Bench~\citep{yuan2025cg} further examines indirect-call reasoning by asking models to resolve function-pointer targets from C/C++ source contexts. 
Together, these benchmarks demonstrate the value of call relations as a diagnostic target beyond end-task correctness.
However, they focus on small instances with few functions and pre-selected contexts, reducing the need to discover and integrate evidence across large program structures.
They also primarily evaluate LLMs under model-only settings, without systematically studying how tool access or external information affects reasoning.
Moreover, both operate at the source-code level, where high-level cues such as types, variable names, and pointer assignments remain available. 
Our work extends this direction to stripped binaries, where such information is unavailable and indirect-call targets are not explicitly represented at callsites. 
We further scale evaluation from small function sets to whole programs, with an average of 941 functions per instance, and broaden the evaluation regime to include agents.
\section{\sys Benchmark}
\subsection{Preliminary}

\PP{Definition of Indirect Calls}
An indirect call invokes a function whose target address is determined at runtime rather than statically encoded at compile time. 
At the binary level, the call target is typically stored in a register or memory location and then invoked indirectly.
The task of binary indirect-call target recovery is to identify all possible targets of an indirect callsite in a binary while minimizing the inclusion of infeasible targets.

\PP{Existing Works on Binary Indirect-call Recovery}
Binary indirect-call target recovery remains challenging for existing static analysis techniques. 
Current methods often face a trade-off between soundness and precision: some~\citep{gao2025bindsa} miss valid targets, which is undesirable for downstream analyses such as control-flow integrity and taint analysis, while others~\citep{van2016tough, kim2021refining} conservatively report many infeasible targets and impose a substantial burden on subsequent analysis. 
This gap motivates binary indirect-call target recovery as our core task, requiring LLM-based systems to recover valid targets while filtering infeasible ones.

\PP{Indirect-call Recovery with Interprocedural Reasoning}
Interprocedural reasoning, where interprocedural refers to operations across different functions in software \citep{muchnick1997advanced}, means understanding interprocedural semantics and recovering cross-function relations.  
Indirect-calls form a natural such relation.
Precisely recovering their targets from binaries requires examining potential target functions and reasoning about their feasibility. 
Local analysis alone may therefore be imprecise, making indirect-call recovery suited for probing LLMs’ interprocedural reasoning on binaries.

\subsection{Benchmark Construction}
\label{sec:benchmark_construction}

\begin{figure*}[t]
    \centering
    \includegraphics[width=\textwidth]{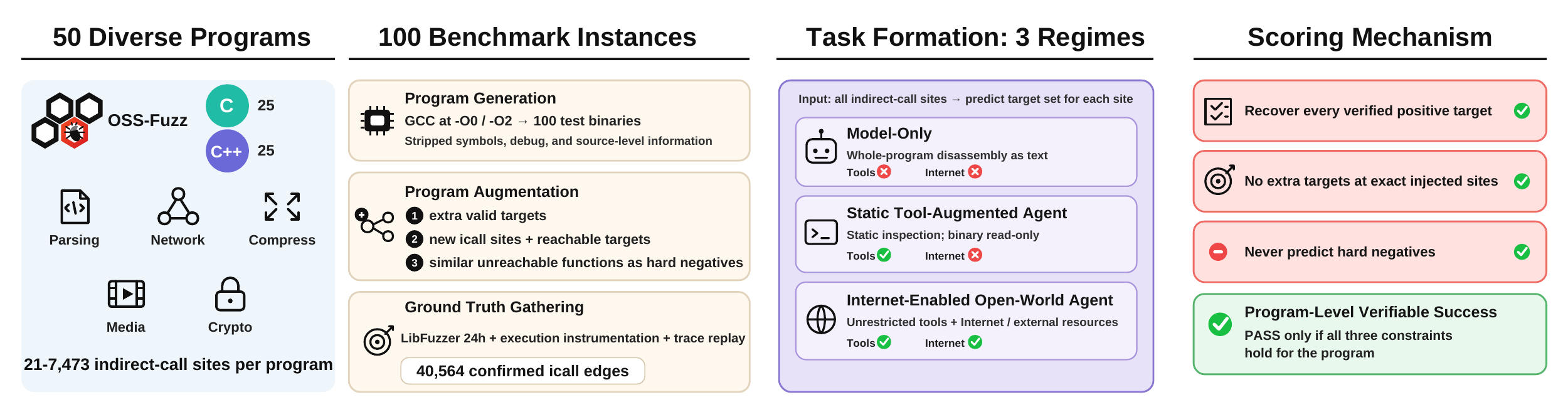}
    \caption{Overview of \sys Construction}
    \label{fig:system_overview}
\end{figure*}

\autoref{fig:system_overview} provides an overview of \sys construction, which we introduce in detail as follows.

\PP{Program Selection and Generation}
We select 50 real-world programs (\autoref{app:benchmark_list}) from OSS-Fuzz \citep{ossfuzz}, Google's fuzzing infrastructure for critical open-source projects, comprising 25 C and 25 C++ programs. 
OSS-Fuzz also provides a mature fuzzing infrastructure and large corpora that support our ground-truth collection process.
The selected programs span diverse application domains, including parsing, networking, compression, multimedia, and cryptography. 
They also vary substantially in complexity and indirect-call density, ranging from 27 to 7,991 functions and from 3,050 to 1,034,208 lines of code, with 21 to 7,473 indirect-call sites.
To further evaluate robustness to compiler optimization, we compile each program with GCC 13.3.0~\citep{gcc} at \texttt{-O0} and \texttt{-O2}, yielding 100 test binaries.
The resulting binaries are stripped to remove static symbols and debug information.
Optimization can alter function boundaries, and indirect-call structures, including introducing or eliminating indirect calls. 
This setup tests whether models can consistently recover interprocedural relations across different binary representations.

\PP{Program Augmentation}
To complement the naturally occurring indirect-call relations, we introduce three types of source-level injections. 
The primary goal is to increase the amount and diversity of ground-truth callsite--target relations available for evaluation, particularly positive relations that can be unambiguously verified. 
These injected relations are also absent from public versions of the programs, providing an additional safeguard against reliance on memorized program structures.
Specifically, we introduce: (1) additional valid targets for existing indirect-call sites, expanding the set of positive callsite--target relations; (2) new indirect-call sites along real execution paths at varying depths, together with candidate targets constructed such that the complete feasible target set is known by construction, providing an exact-precision subset for evaluation; and (3) functions that are similar to valid targets but are never invoked by any indirect-call sites, providing hard negatives to test whether models recover actual callsite--target relations rather than relying solely on function-level similarity.
All injected targets from Steps 1 and 2 are verified through real execution, whereas Step 3 targets are unreachable by design.

\PP{Ground Truth Gathering}
Following established practice in prior static indirect-call analysis work \cite{kim2021refining, gao2025bindsa, du2026sok}, we construct the ground truth via dynamic profiling by executing each program on a large corpus of fuzz-generated test cases and recording the resulting indirect callsite--target edges.
We fuzz each program with LibFuzzer~\citep{libfuzzer} for 24 hours on a 96-core CPU machine, using OSS-Fuzz corpora as seeds.
All collected edges are replayed with their corresponding inputs to confirm reproducibility.
In total, this process yields 40,564 confirmed indirect-call edges.
Dynamic profiling is inherently incomplete and therefore provides only a lower bound on the true target set. 
However, every observed target is a verified call relation that any sound recovery method must recover; missing any such target is therefore an unambiguous error. 
We use these verified relations as minimum correctness requirements, which motivates the program-level verifiable success criterion defined in \autoref{sec:scoring}.

\PP{Task Formulation}
We first extract all indirect-call sites from each binary and represent them as a list of instruction addresses. 
Based on the three analysis regimes described in \autoref{sec:intro}, we formulate the tasks as follows. 
In the \textbf{model-only} regime, we disassemble the entire binary and collect the assembly code together with the necessary data and linkage sections into a textual representation, which is provided to the LLM along with the indirect-call site list; the model must identify the target set for each site without access to external tools, execution, or the Internet. 
In the \textbf{static tool-augmented agent} regime, we provide a container equipped with \texttt{objdump}~\citep{objdump}, \texttt{radare2}~\citep{radare2} and other standard binary-inspection tools, while restricting network access to the LLM API and making the evaluated binary read-only. 
As this setting reflects a typical static-analysis workflow, allowing the agent to inspect the binary using available tools, but not to execute or modify it, we use it as the default evaluation setting.
In the \textbf{Internet-enabled open-world agent} regime, the agent operates in an unrestricted container where it may install additional tools, access the Internet for documentation or external resources, and execute the binary for dynamic analysis and validation, thereby allowing it to combine static analysis, dynamic execution, and external information to complete the task. 
The corresponding task prompts are shown in \autoref{app:prompt}.

\subsection{Scoring Mechanism}
\label{sec:scoring}

We score each program against the ground truth introduced in \autoref{sec:benchmark_construction}.
For program instance $p$, let $\mathcal{E}_p$ denote the callsites with available ground-truth targets, where $G_i^{+}$ is the set of verified targets for callsite $i$; let $\mathcal{X}_p \subseteq \mathcal{E}_p$ denote the injected exact callsites from Program Augmentation Step~2, where $G_i^{e}$ is the corresponding exact target set; and let $\mathcal{N}_p$ denote the injected known non-targets from Program Augmentation Step~3 that must not be predicted as valid indirect-call targets. The system under evaluation predicts a target set $\hat{G}_i$ for each callsite $i$.

We treat the available ground truth as the minimum correctness requirement. 
All verified targets must be recovered, injected exact callsites must contain no additional targets, and injected known non-targets must never be predicted:
\begin{equation}
    G_i^{+} \subseteq \hat{G}_i \;\; (i\in\mathcal{E}_p), \qquad
    \hat{G}_i \setminus G_i^{e}=\emptyset \;\; (i\in\mathcal{X}_p), \qquad
    \hat{\mathcal{T}}_p\cap\mathcal{N}_p=\emptyset,
\end{equation}
where $\hat{\mathcal{T}}p=\bigcup{i\in\mathcal{C}_p}\hat{G}_i$ is the union of all predicted targets over the set $\mathcal{C}_p$ of indirect callsites in program $p$.

\paragraph{Program-Level Verifiable Success Rate.}
Following the minimum-correctness requirement above, a program is considered a verifiable success only if it contains no verifiable error under the available ground truth:
\begin{equation}
    \mathrm{VS}(p)=\mathbb{I}\!\left[(\forall i\in\mathcal{E}_p,\;G_i^{+}\subseteq\hat{G}_i)\land(\forall i\in\mathcal{X}_p,\;\hat{G}_i\setminus G_i^{e}=\emptyset)\land(\hat{\mathcal{T}}_p\cap\mathcal{N}_p=\emptyset)\right].
\end{equation}
We report the \emph{Program-Level Verifiable Success Rate} (PVSR):
\begin{equation}
    \mathrm{PVSR}=\frac{1}{|\mathcal{P}|}\sum_{p\in\mathcal{P}}\mathrm{VS}(p).
\end{equation}
PVSR is our primary benchmark metric and measures the fraction of programs with no verifiable error under the available ground truth.
Standard precision and recall are not directly comparable because task failures may prevent a system from producing predictions for all requested callsites, leading to incomplete outputs and inconsistent evaluation sets across systems.
Since the verified targets form only a lower bound on the true target sets, PVSR is an optimistic upper bound on true program-level correctness: additional ground truth may reveal further failures, but cannot invalidate observed ones. 
We further analyze detailed failure statistics and their underlying causes in \autoref{sec:failure}.

\section{Evaluation}
\subsection{Experimental Setup}
\label{sec:eval_setup}

\PP{Models and Agents}
We evaluate five recent LLMs: \textsc{GPT-5.6 Sol}~\citep{openai2026gpt56sol},
\textsc{GPT-5.6 Luna}~\citep{openai2026gpt56luna},
\textsc{Claude Opus 5}~\citep{anthropic2026claudeopus5},
\textsc{GLM-5.3 Flash}~\citep{zai2026glm53flash}, and
\textsc{DeepSeek-V4.1 Flash}~\citep{deepseek2026v41flash}.
For agent-based analysis regimes, we use the native agent harness provided for GPT (Codex~\citep{openai2025codex}), Claude (Claude Code~\citep{anthropic2025claudecode}), and GLM (ZCode~\citep{zai2026zcode}), while DeepSeek is evaluated through OpenCode~\citep{opencode2025}.
All models use the max reasoning effort and are accessed via their official APIs.

\PP{Run Protocol}
For both agent-based regimes, each agent runs in a container with 4 CPU cores and 8 GB of memory, with a limit of three hours or 600 agent turns per run, whichever is reached first. 
We perform one run for each model, regime, and program instance under the same resource budget. 
We evaluate single-run performance rather than best-of-$k$ or pass@$k$, as selecting the best of $k$ runs is impractical in real-world indirect-call recovery where ground truth is unavailable.

\subsection{How Capable Are LLM Agents in Binary Interprocedural Reasoning?}
\label{sec:rq1}

\begin{table}[t]
    \centering
    \small
    \setlength{\tabcolsep}{6pt}
    \caption{Overall performance under the default static tool-augmented regime.}
    \label{tab:overall_results}
    \begin{tabular}{llcccc}
        \toprule
        \textbf{Model} &
        \textbf{Agent} &
        \textbf{PVSR $\uparrow$} &
        \textbf{Avg. Time} &
        \textbf{Avg. Turns} &
        \textbf{Avg. Cost} \\
        \midrule
        Claude Opus 5       & Claude Code & \textbf{81\%} & 33m  & 93  & \$12.27 \\
        GPT-5.6 Sol         & Codex       & 72\%          & 43m  & 172 & \$12.48 \\
        DeepSeek-V4.1 Flash & OpenCode    & 46\%          & 41m  & 138 & \$0.24  \\
        GLM-5.3 Flash       & ZCode       & 25\%          & 111m & 102 & \$0.58  \\
        GPT-5.6 Luna        & Codex       & 23\%          & 94m  & 330 & \$0.81  \\
        \bottomrule
    \end{tabular}
\end{table}

\PP{Overall Performance}
\textit{Current LLM-based agents show substantial capability gaps, and even the strongest system exhibits frequent verifiable failures.}

We first evaluate LLM-based agents under the static tool-augmented regime, our default setting.
Agents may actively inspect target binaries using static-analysis tools, but cannot execute them or access external information, reflecting a typical static-analysis workflow. 

As shown in \autoref{tab:overall_results}, PVSR varies substantially across systems, ranging from 23\% to 81\%.
Claude Opus 5 with Claude Code achieves the highest PVSR at 81\%, followed by GPT-5.6 Sol with Codex at 72\%.
DeepSeek-V4.1 Flash reaches 46\%, while GLM-5.3 Flash and GPT-5.6 Luna achieve 25\% and 23\%, respectively.
This wide spread shows that \sys effectively differentiates the interprocedural reasoning capabilities of current LLM-based systems.

Importantly, these numbers are optimistic by construction: PVSR only requires the absence of errors verifiable under the available ground truth.
Thus, even the strongest system exhibits verifiable failures on 19\% of program instances, spanning 742 incorrect callsites, while additional unobserved targets could only reveal further failures.
These results indicate substantial remaining headroom even at the top end of performance.
We examine the underlying failure patterns in detail in \autoref{sec:failure}.

The results further suggest that these performance differences are not simply explained by greater resource consumption. 
Longer runtimes or more interaction turns do not consistently yield higher PVSR. 
For example, GPT-5.6 Luna uses substantially more turns than the two strongest systems while achieving a much lower PVSR. 
This indicates that the observed capability gap reflects differences in analysis effectiveness rather than interaction budget alone.

\PP{Impact of Program Characteristics}
\textit{Optimization has relatively little impact on performance.}

We next examine whether this capability varies with three program characteristics: indirect-call scale, compiler optimization, and programming language.
For indirect-call scale, we divide the 50 programs based on the median number of indirect-call sites, yielding small- and large-scale groups.

\begin{figure*}[t]
    \centering
    \includegraphics[width=\textwidth]{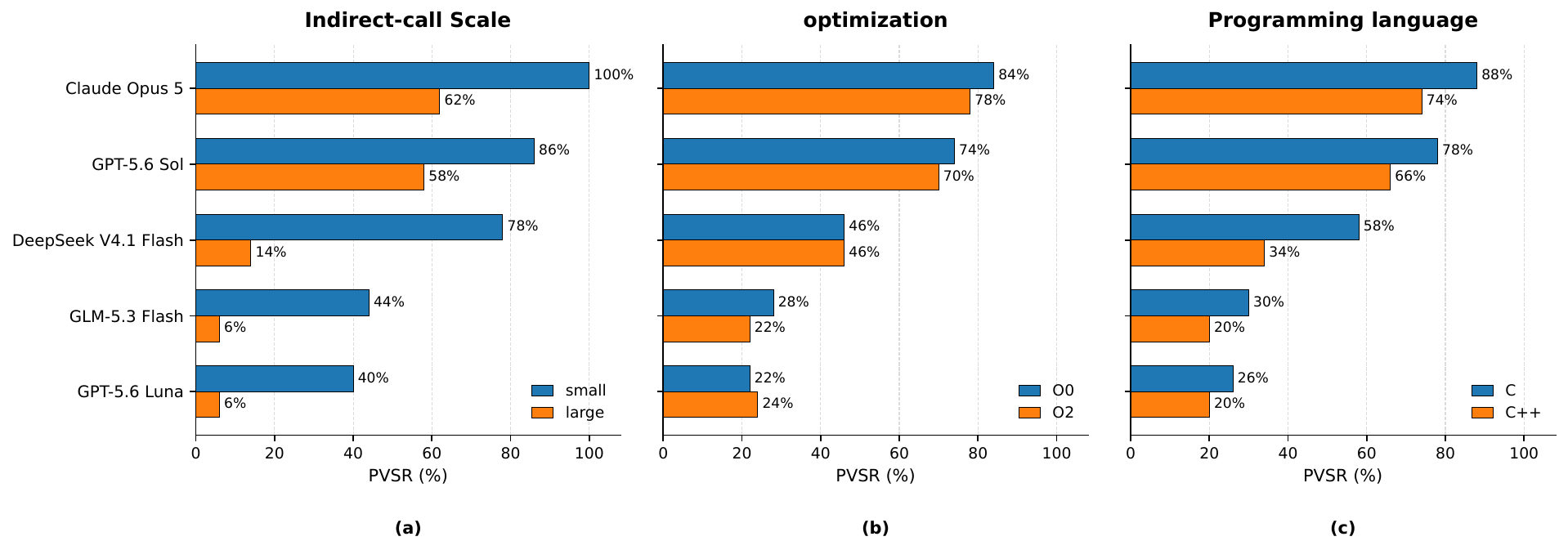}
    \caption{Impact of program characteristics on PVSR under the
    static tool-augmented regime.}
    \label{fig:rq1_factors}
\end{figure*}

As shown in \autoref{fig:rq1_factors}, performance varies most strongly with indirect-call scale. 
From small- to large-scale programs, PVSR drops substantially across all five systems, which is intuitive: as agents need to recover more indirect calls, they have more opportunities to make at least one error. 
C programs also consistently achieve higher PVSR than C++ programs, with gaps ranging from 6 to 24 percentage points, although this comparison is partly coupled with indirect-call scale: among the 25 small-scale programs, 15 are C and 10 are C++, whereas among the 25 large-scale programs, 10 are C and 15 are C++. 
In contrast, compiler optimization has only a limited impact, with O0 and O2 differing by at most 6 percentage points for four systems and showing no difference for DeepSeek V4.1 Flash. 
This suggests that current LLM agents are relatively insensitive to variations in binary representation introduced by different optimization levels.

\subsection{How Do Different Analysis Regimes Affect Performance?}

\label{sec:rq2}

\begin{figure*}[t]
    \centering
    \includegraphics[width=\textwidth]{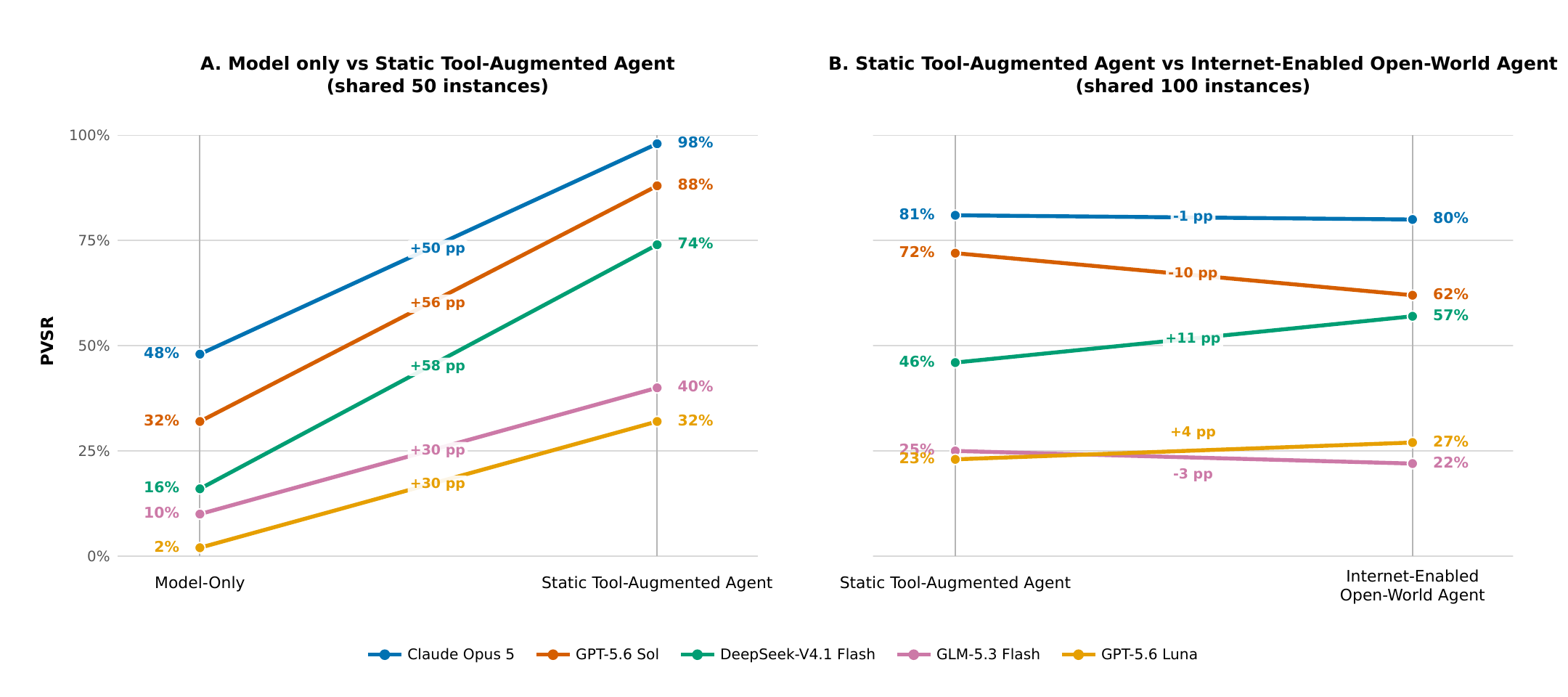}
    \caption{Side-by-side Change Comparison for the three analysis regimes. Panel A is evaluated on the 50 context-eligible instances, while Panel B is evaluated on the full set of 100 instances.}
    \label{fig:analysis_regime}
\end{figure*}

\PP{Model-Only vs.\ Static Tool-Augmented Agent}
\textit{Tool augmentation consistently improves performance by reducing the reasoning burden imposed by large textual contexts.}

Model-only analysis requires the complete program representation to fit within the model context. 
We therefore evaluate both model-only and static tool-augmented settings on the same subset of 50 program instances whose tiktoken tokenizer \citep{tiktoken} representations contain fewer than one million tokens. 
Any run that fails to produce a valid final prediction is counted as failure, whether due to context limitations, or unsuccessful reasoning.
We treat context-limit failures as genuine failures because the available context window is itself an important aspect of an LLM model's capability.
A detailed breakdown is provided in ~\autoref{app:failure_break_down}.

As shown in \autoref{fig:analysis_regime}, static tool-augmented agent improves PVSR by 30--58 percentage points across all evaluated models. 
The largest gains occur for DeepSeek-V4.1 Flash and GPT-5.6 Sol, increasing from 16\% to 74\% and from 32\% to 88\%, respectively. 
Inspection of agent trajectories shows a consistent strategy: agents iteratively use tools to inspect relevant functions, reason over dispatcher and call-site structure, and track pointer values across functions. 
This stepwise navigation avoids reasoning over a monolithic, long-context program representation and instead decomposes the task into manageable subproblems. 
In contrast, models’ intrinsic reasoning remains substantially limited.
Failures become increasingly frequent as program size and context length grow, and even the best-performing model achieves a pass rate below 50\% in the model-only setting.

\PP{Static Tool-Augmented Agent vs.\ Internet-Enabled Open-World Agent}
\textit{Broader access is not uniformly beneficial; its value depends on whether agents can ground additional evidence in binary-specific interprocedural relations.}

We next compare the static tool-augmented and Internet-enabled open-world agents across all 100 program instances. 
In contrast to the consistent gains observed with static tool augmentation, moving to the Internet-enabled open-world regime has a heterogeneous effect. 
DeepSeek-V4.1 Flash improves by 11 percentage points and GPT-5.6 Luna by 4 points, while Claude Opus 5 remains nearly unchanged. 
In contrast, GPT-5.6 Sol and GLM-5.3 Flash decrease by 10 and 3 points, respectively. 
Thus, broader access to execution, tools, and external information can benefit some systems, but does not uniformly translate into better interprocedural reasoning performance.

A closer examination of the trajectories reveals substantial churn beneath these aggregate results: every agent both gains and loses passed instances in the open-world setting (\autoref{app:score_change}). 
Internet access can help by providing source code and documentation that expose relevant targets and reduce the need to reconstruct program structure from the binary.
However, external artifacts can also distract agents from binary-specific evidence. 
In many GPT-5.6 Sol trajectories, agents spend more effort rebuilding or interpreting source code and less on callsite structure and dispatcher logic.
Moreover, mismatches between the retrieved source and the analyzed binary can lead to incorrect conclusions, as seen in several Claude Opus 5 trajectories.
These failures further highlight the value of program augmentation in \autoref{sec:benchmark_construction}, which provides a controlled setting that requires reasoning about the actual binary rather than relying solely on external artifacts.
Additional execution privileges introduce a similar trade-off. 
They help agents validate candidate targets, but can also make them overly conservative, causing them to report only dynamically confirmed targets and miss valid ground-truth ones. 
These effects differ across systems: DeepSeek-V4.1 Flash benefits substantially from execution-based confirmation, whereas GPT-5.6 Sol suffers most from reduced direct binary analysis.

Overall, broader access is not inherently beneficial. 
Its effectiveness depends on how well the agent uses external evidence and additional privileges to understand the binary’s interprocedural semantics. 
Moreover, even under this strongest setting, the best-performing agent still makes substantial errors, indicating that our benchmark is far from saturated.

\subsection{Failure Analysis}
\label{sec:failure}

We further examine why current systems fail to recover indirect-call targets and, conversely, what enables individual systems to solve cases that others fail.

\PP{Common Reasoning Failures}
We first examine the cases where all agents failed and identify three recurring reasoning problems. 
First, agents often misidentify where a function pointer comes from. 
For example, a callback or global function-pointer variable defined inside the binary may be mistaken for an external reference, causing the agent to miss a valid in-binary target. 
Second, an agent may correctly recognize the dispatch pattern, such as a vtable, handler table, or callback table, but associate it with the wrong object. 
As a result, the predicted targets look structurally reasonable but belong to the wrong class, table, or structure field. 
Third, some difficult cases require reasoning beyond the local functions, where valid targets are revealed only through related functions, static initialization, and control-flow reconvergence.
For instance, the relevant value may be created in one function and used at the indirect callsite in another.
Detailed case studies illustrating these failure modes are provided in \autoref{app:reasoning-failure-cases}.

\PP{Unique Pass Cases}
We next inspect difficult instances passed by exactly one system, with Claude Opus 5 accounting for 11 and GPT-5.6 Sol for 6.
These cases reveal several complementary reasoning strategies that enable one system to recover relations missed by the others.
Claude Opus 5 explicitly models the relation between function-pointer assignments, structure fields, and indirect calls, allowing it to recover targets across functions accurately. 
GPT-5.6 Sol identifies the relevant callback relation and resolves its members exhaustively before committing, which helps maintain high precision.
Detailed case studies illustrating these
strategies are provided in \autoref{app:successful-reasoning-strategies}.

\PP{Takeaway}
Current agents struggle to determine function-pointer provenance, bind dispatch structures to the correct semantic object, and propagate information across function boundaries.
The strongest successes arise when agents preserve these relations through explicit cross-function abstractions and exhaustive relation-specific enumeration.

\section{Discussion}
\PP{Scope of the Benchmark}
\sys focuses on stripped x86-64 Linux ELF binaries compiled from C and C++ programs.
While this setting is widely adopted in existing work on binary indirect call recovery, the current benchmark does not encompass the additional challenges posed by other architectures, platforms, binary formats, programming languages, or a broader range of compilers and toolchains.
Extending the benchmark along these dimensions is a direction of future work.

\PP{Evaluation Scope}
Indirect-call recovery provides a concrete probe of interprocedural reasoning, but does not capture all forms of program understanding.
Moreover, agent-based results reflect both the underlying model and its harness. 
We provide an ablation study in \autoref{app:agent_scaffold} to examine the effect of the harness.
Therefore, our agent results should be interpreted as system-level performance on this specific reasoning task rather than a complete measure of model capability.


\PP{Ground-Truth Coverage and Robustness} Ground truth for naturally occurring indirect calls is collected through dynamic profiling and therefore represents verified positives rather than exhaustive target sets. 
Across the benchmark, dynamic profiling yields 40,564 confirmed edges and is largely saturated, with fuzzing beyond 18 hours producing only marginal growth in the verified relation set. 
To guard against over-prediction under incomplete ground truth, we additionally inject 802 callsites with exact target sets known by construction, together with unreachable hard negatives, as described in \autoref{sec:benchmark_construction}.
Under this criterion, agents must recover the exact target sets for the injected callsites to pass. 
We find no evidence that high-performing models succeed by indiscriminately expanding target sets; instead, their failures more often stem from missing valid targets than from overpredicting infeasible ones.
A detailed breakdown of errors under each constraint is provided in \autoref{app:failure_break_down}. 
As a further sanity check, we evaluate the state-of-the-art static-analysis baseline BinDSA~\citep{gao2025bindsa}, which achieves 0\% PVSR on our benchmark and satisfies all applicable correctness constraints on only 6.32\% of ground-truth-covered sites, indicating that the benchmark is not trivially solved by existing indirect-call analysis. 
Taken together, these results do not establish exhaustive ground truth, but suggest that residual incompleteness is unlikely to dominate the benchmark conclusions, while the poor performance of existing static analysis further highlights the difficulty of the benchmark.

\section{Conclusion}
We introduce \sys, a benchmark for evaluating interprocedural reasoning in binary programs through indirect-call target recovery. 
Our results show that current LLM agents still struggle with complex cross-function reasoning, and broader external information and access do not consistently improve performance.
Overall, the findings suggest that future binary-analysis agents need stronger representations of cross-function semantics, especially for tracking pointer provenance, structure fields, and non-local control flow, while also learning to effectively use and ground external information and expanded access in the specific binary.

\clearpage

\subsection*{Ethics statement}
\sys evaluates interprocedural reasoning through indirect-call target recovery in open-source binary programs, and the benchmark tasks themselves do not involve exploit development, vulnerability weaponization, or other harmful actions. 
The evaluated programs are drawn from open-source projects, and the benchmark is designed to measure program-analysis capabilities rather than offensive security behavior.
For the Internet-enabled open-world evaluation, where agents were allowed broader tool and Internet access, we manually reviewed all agent trajectories and access histories to verify that the evaluated systems did not perform harmful or unintended actions. 
The experiments were conducted solely for research evaluation, and we found no harmful activity during these runs.

\subsection*{Reproducibility statement}
The task formulation, analysis regimes, prompts, and evaluation protocol are specified in the main paper and appendix. 
We will release all benchmark binaries used in the evaluation through an anonymous repository (https://anonymous.4open.science/r/icall-bench/). 
To preserve the validity of the Internet-enabled open-world setting, the test ground truth, including dynamically verified call relations, injected exact target sets, and hard negatives, will be kept private. 
Public release of these labels, or a publicly accessible verification service, could compromise the benchmark by allowing evaluated open-world agents to directly retrieve, query, or adaptively infer information about the hidden test outcomes. 
After the review period, we will provide a controlled submission channel through which predictions on the released binaries can be evaluated against the private ground truth, with aggregate evaluation results returned to submitters.

\clearpage

\bibliography{iclr2027_conference}
\bibliographystyle{iclr2027_conference}

\appendix
\clearpage

\section{Benchmark Program List}
\label{app:benchmark_list}
\begin{table*}[!htbp]
\centering
\caption{
Benchmark programs used in \sys, including their programming languages,
versions, lines of code, and number of functions.
Programming language information is sourced from OSS-Fuzz \citep{ossfuzz}.  
}
\label{tab:benchmark-programs}

\scriptsize

\begin{minipage}[t]{0.49\textwidth}
\centering
\resizebox{\linewidth}{!}{%
\begin{tabular}{@{}lclrr@{}}
\toprule
\textbf{Program} & \textbf{Lang.} & \textbf{Version}
& \textbf{LOC} & \textbf{Funcs} \\
\midrule
\prog{brotli}      & C   & v1.1.0              & 14,168    & 71    \\
\prog{bzip2}       & C   & bzip2-1.0.8         & 5,236     & 54    \\
\prog{c-ares}      & C   & v1.34.5             & 45,538    & 697   \\
\prog{capstone}    & C   & 5.0.9               & 1,034,208 & 1,301 \\
\prog{civetweb}    & C   & v1.16               & 27,077    & 173   \\
\prog{flac}        & C++ & 1.5.0               & 44,494    & 899   \\
\prog{harfbuzz}    & C++ & 10.4.0              & 138,598   & 1,644 \\
\prog{hiredis}     & C   & v1.2.0              & 7,259     & 183   \\
\prog{htslib}      & C   & 1.23                & 97,650    & 1,337 \\
\prog{http-parser} & C   & (untagged)          & 3,050     & 27    \\
\prog{jbig2dec}    & C++ & 0.20                & 16,014    & 140   \\
\prog{jsonnet}     & C++ & v0.20.0             & 55,593    & 1,311 \\
\prog{lcms}        & C   & lcms2.17            & 41,312    & 1,011 \\
\prog{libarchive}  & C++ & v3.8.1              & 128,501   & 1,509 \\
\prog{libcbor}     & C++ & v0.12.0             & 6,461     & 259   \\
\prog{libevent}    & C++ & release-2.2.2-alpha & 53,506    & 958   \\
\prog{libfdk-aac}  & C++ & v2.0.3              & 138,333   & 940   \\
\prog{libgd}       & C++ & gd-2.3.3            & 46,717    & 339   \\
\prog{libhtp}      & C   & 0.5.50              & 24,802    & 455   \\
\prog{libical}     & C++ & v3.0.20             & 48,282    & 1,306 \\
\prog{libpcap}     & C   & libpcap-1.10.5      & 53,022    & 437   \\
\prog{libpng}      & C++ & v1.6.58             & 38,417    & 486   \\
\prog{libraw}      & C++ & 0.21.5              & 55,159    & 728   \\
\prog{libssh2}     & C++ & libssh2-1.11.1      & 41,740    & 429   \\
\prog{libtiff}     & C++ & v4.7.2              & 54,144    & 608   \\
\bottomrule
\end{tabular}%
}
\end{minipage}
\hfill
%
\begin{minipage}[t]{0.49\textwidth}
\centering
\resizebox{\linewidth}{!}{%
\begin{tabular}{@{}lclrr@{}}
\toprule
\textbf{Program} & \textbf{Lang.} & \textbf{Version}
& \textbf{LOC} & \textbf{Funcs} \\
\midrule
\prog{libucl}      & C   & 0.9.2                  & 18,593  & 302   \\
\prog{libwebp}     & C   & v1.3.2                 & 74,216  & 1,009 \\
\prog{libxml2}     & C   & v2.13.8                & 192,230 & 2,477 \\
\prog{lz4}         & C   & v1.10.0                & 11,389  & 192   \\
\prog{mbedtls}     & C   & v3.6.7                 & 202,100 & 1,615 \\
\prog{miniz}       & C   & 3.0.2                  & 9,350   & 151   \\
\prog{mongoose}    & C   & 7.19                   & 30,187  & 288   \\
\prog{njs}         & C   & 0.9.9                  & 87,005  & 1,181 \\
\prog{openh264}    & C++ & v2.6.0                 & 42,078  & 577   \\
\prog{openjpeg}    & C++ & v2.5.4                 & 48,424  & 464   \\
\prog{opus}        & C   & v1.5.2                 & 57,654  & 385   \\
\prog{pcre2}       & C   & pcre2-10.45            & 66,298  & 171   \\
\prog{simdutf}     & C++ & v5.6.4                 & 34,200  & 931   \\
\prog{snappy}      & C++ & 1.2.2                  & 4,560   & 108   \\
\prog{spirv-cross} & C++ & vulkan-sdk-1.4.313.0   & 78,309  & 3,931 \\
\prog{sql-parser}  & C++ & (untagged)             & 14,782  & 335   \\
\prog{tidy-html5}  & C   & 5.8.0                  & 57,189  & 763   \\
\prog{uriparser}   & C++ & uriparser-1.0.2        & 14,018  & 295   \\
\prog{wabt}        & C++ & 1.0.36                 & 57,174  & 4,118 \\
\prog{wasm3}       & C   & v0.5.0                 & 16,388  & 824   \\
\prog{wavpack}     & C++ & 5.9.0                  & 19,832  & 210   \\
\prog{xerces-c}    & C++ & v3.2.5                 & 249,870 & 7,991 \\
\prog{xz}          & C   & v5.8.1                 & 38,561  & 378   \\
\prog{yara}        & C++ & v4.5.2                 & 70,276  & 820   \\
\prog{zstd}        & C   & v1.5.7                 & 23,350  & 211   \\
\bottomrule
\end{tabular}%
}
\end{minipage}

\end{table*}

\clearpage
\section{Detailed Evaluation Setup and Prompt for the Three Analysis Regimes}
\label{app:prompt}

All our evaluations were conducted on a machine equipped with an AMD EPYC 9654 96-core processor and 768 GB of memory. Detailed prompts for each task are provided below.

\begin{figure*}[!htbp]
\centering

\begin{minipage}{\textwidth}
\begin{promptbox}
\begin{lstlisting}[style=promptstyle]
Your task is to find all the possible targets of each indirect call site.

This file is the whole task. It has three sections, in this order, each opened by a line beginning `== ` under a rule of `=`:
    CALL SITES      the addresses you must answer for, one per line
    DATA            the program's ELF tables and data sections
    DISASSEMBLY     every executable section, disassembled end to end

Everything below this section was derived from one stripped ELF executable and nothing else. That program has no symbol table, so nothing in it has a name: functions, call sites and data are identified by address alone, written in lowercase hex. Every address here is a link-time virtual address of the ELF image -- DATA states whether that image is position-independent, and if it is, read them with a load bias of zero. The addresses under CALL SITES are in the same space as every other address in this file.

## What to produce

For every address listed under CALL SITES, give the set of functions IN THIS PROGRAM that the instruction at that address can transfer control to at run time. Answer with a single JSON object and nothing else:

    {"0x1014": ["0x5f24", "0x6110"], "0x10ff": ["0x9460"], "0x112f": []}

* one key per CALL SITES line, spelled exactly as that line spells it;
* each value a list of target addresses, lowercase `0x` hex, deduplicated and ascending, every one of them the ENTRY address of a function of this program.

==================================================================
== CALL SITES
==================================================================

...

==================================================================
== DATA
==================================================================

...

==================================================================
== DISASSEMBLY
==================================================================

...
\end{lstlisting}
\end{promptbox}
\end{minipage}

\caption{Prompt for Model-Only Analysis Regime.}
\label{fig:model_only_prompt}

\end{figure*}

\begin{figure*}[!htbp]
\centering

\begin{minipage}{0.99\textwidth}
\begin{promptbox}
\begin{lstlisting}[style=promptstyle]
`bin` in the work directory is a binary executable.
`site.txt` lists the addresses of indirect call sites in that binary, one per line, as lowercase hexadecimal with an `0x` prefix (for example `0x4011a0`).

Your task is to find all the possible targets of each indirect call site. Resolve them by statically analyzing the binary.

## Tooling constraints

These are hard rules for this task.

- **Static analysis only - never execute `bin`.** Do not run it directly, do not run it under a debugger, tracer or emulator (`gdb`, `strace`, `ltrace`), do not load it into a process (`dlopen`, `LD_PRELOAD`), and do not execute pieces of its code from a harness of your own. Every target you report has to follow from static analysis of the binary, not from observing its execution.

- **Use static-analysis tools available in this container.** You may use any static binary-analysis tools, disassemblers, decompilers, binary-analysis frameworks, and supporting command-line tools that are already available in the container. You may combine multiple tools and write your own scripts.

- **You have no internet access.** The container firewall rejects every outbound destination except the endpoint your own model provider is served from. `apt-get install`, `pip install`, `git clone` and `curl` against any other host will fail.

## Output

Every address in this task - those in `site.txt`, and both the keys and the values of `result.json` - is a virtual address in the binary's own address space, written as lowercase hex with an `0x` prefix and no zero padding.

Produce two files in the work directory.

1. `result.json` - a JSON object with the following structure:

   {
     "0x<site_address>": ["0x<target_address1>", "0x<target_address2>", ...],
     ...
   }

   Each key is an indirect call site address from `site.txt`, and the value is a list of all possible target addresses for that call site.

   - Include an entry for every address in `site.txt`, and for no other address.
   - Reproduce each key exactly as it appears in `site.txt`.
   - A site whose targets you cannot resolve gets an empty list `[]` rather than being left out.
   - List each target once, sorted ascending by address.

2. `report.md` - a short written summary that describes the analysis performed, the number of indirect call sites analyzed, and any notable findings or patterns observed in the indirect call targets. Include which static-analysis tools and analysis techniques you used, and briefly explain how you went from the binary to the final target sets.
\end{lstlisting}
\end{promptbox}
\end{minipage}

\caption{Prompt for Static Tool-Augmented Agent.}
\label{fig:static-agent-prompt}

\end{figure*}

\begin{figure*}[!htbp]
    \centering

    \begin{minipage}{0.99\textwidth}
        \begin{promptbox}
\begin{lstlisting}[style=promptstyle]
`bin` in the work directory is a binary executable.
`site.txt` lists the addresses of indirect call sites in that binary, one per line, as lowercase hexadecimal with an `0x` prefix (for example `0x4011a0`).

Your task is to find all the possible targets of each indirect call site.

Every address in this task - those in `site.txt`, and both the keys and the values of `result.json` - is a virtual address in the binary's own address space, written as lowercase hex with an `0x` prefix and no zero padding.

Produce two files in the work directory.

1. `result.json` - a JSON object with the following structure:

   {
     "0x<site_address>": ["0x<target_address1>", "0x<target_address2>", ...],
     ...
   }

   Each key is an indirect call site address from `site.txt`, and the value is a list of all possible target addresses for that call site.

   - Include an entry for every address in `site.txt`, and for no other address.
   - Reproduce each key exactly as it appears in `site.txt`.
   - A site whose targets you cannot resolve gets an empty list `[]` rather than being left out.
   - List each target once, sorted ascending by address.

2. `report.md` - a short written summary that describes the analysis performed, the number of indirect call sites analyzed, and any notable findings or patterns observed in the indirect call targets.

Use whatever tooling you like; you can install any tools you want.
You have open internet access and may download any resources you find helpful.
\end{lstlisting}
        \end{promptbox}
    \end{minipage}

    \caption{Prompt for Internet-Enabled Open-World Agent.}
    \label{fig:open_world_prompt}
\end{figure*}

\clearpage
\section{Detailed Breakdown of the Result of the Three Analysis Regimes}
\label{app:failure_break_down}

\begin{table*}[!htbp]
\centering
\small
\renewcommand{\arraystretch}{1.15}

{\setlength{\tabcolsep}{4pt}
\begin{tabular}{lrrrrrrrr}
\toprule
\multicolumn{9}{c}{\textbf{Model-only}} \\
\midrule
& & \multicolumn{4}{c}{\textbf{Wrong breakdown}} & & \\
\cmidrule(lr){3-6}
Model
& Success
& \shortstack{Miss\\only}
& \shortstack{Extra\\only}
& \shortstack{Miss +\\Extra}
& \shortstack{Decoy\\only}
& Wrong
& \shortstack{Context\\limit}
& Incomplete \\
\midrule
Claude-Opus-5
& 24 & 8  & 0 & 1  & 0 & 9  & 11 & 6 \\

GPT-5.6-Sol
& 16 & 25 & 0 & 1  & 1 & 27 & 7  & 0 \\

DeepSeek-v4.1-Flash
& 8  & 23 & 1 & 11 & 0 & 35 & 4  & 3 \\

GLM-5.3-Flash
& 5  & 21 & 0 & 16 & 0 & 37 & 7  & 1 \\

GPT-5.6-Luna
& 1  & 6  & 0 & 36 & 0 & 42 & 7  & 0 \\
\bottomrule
\end{tabular}
}

\vspace{0.8em}

{\setlength{\tabcolsep}{4pt}
\begin{tabular}{lrrrrrr}
\toprule
\multicolumn{7}{c}{\textbf{Static tool-augmented}} \\
\midrule
& & \multicolumn{3}{c}{\textbf{Wrong breakdown}} & \\
\cmidrule(lr){3-5}
Agent / Model
& Success
& \shortstack{Miss\\only}
& \shortstack{Extra\\only}
& \shortstack{Miss +\\Extra}
& Wrong
& Incomplete \\
\midrule
Claude Code / Claude-Opus-5
& 81 & 13 & 0 & 4  & 17 & 2 \\

Codex / GPT-5.6-Sol
& 72 & 25 & 0 & 0  & 25 & 3 \\

OpenCode / DeepSeek-v4.1-Flash
& 46 & 40 & 1 & 11 & 52 & 2 \\

ZCode / GLM-5.3-Flash
& 25 & 26 & 1 & 17 & 44 & 31 \\

Codex / GPT-5.6-Luna
& 23 & 42 & 1 & 13 & 56 & 21 \\
\bottomrule
\end{tabular}
}

\vspace{0.8em}

{\setlength{\tabcolsep}{4pt}
\begin{tabular}{lrrrrrr}
\toprule
\multicolumn{7}{c}{\textbf{Internet-enabled open-world}} \\
\midrule
& & \multicolumn{3}{c}{\textbf{Wrong breakdown}} & \\
\cmidrule(lr){3-5}
Agent / Model
& Success
& \shortstack{Miss\\only}
& \shortstack{Extra\\only}
& \shortstack{Miss +\\Extra}
& Wrong
& Incomplete \\
\midrule
Claude Code / Claude-Opus-5
& 80 & 15 & 0 & 5  & 20 & 0 \\

Codex / GPT-5.6-Sol
& 62 & 32 & 1 & 5  & 38 & 0 \\

OpenCode / DeepSeek-v4.1-Flash
& 57 & 32 & 1 & 10 & 43 & 0 \\

Codex / GPT-5.6-Luna
& 27 & 42 & 1 & 18 & 61 & 12 \\

ZCode / GLM-5.3-Flash
& 22 & 20 & 1 & 20 & 41 & 37 \\
\bottomrule
\end{tabular}
}

\caption{
Results under three analysis regimes.
Wrong results are partitioned into mutually exclusive error categories.
\textit{Miss only} denotes results that fail to identify one or more
ground-truth targets but do not additionally predict incorrect targets
at the exact callsites.
\textit{Extra only} denotes results that identify all ground-truth targets
but additionally predict one or more incorrect targets at the exact
callsites.
\textit{Miss + Extra} denotes results that both miss at least one
ground-truth target and additionally predict at least one incorrect target
at the exact callsites.
For the model-only setting, \textit{Decoy only} denotes results whose sole
error is the inclusion of a decoy callsite.
No decoy errors occur in the two tool-augmented settings, so the
\textit{Decoy only} column is omitted for those tables.
\textit{Incomplete} indicates that the system failed to produce a complete
result covering all indirect callsites requested in the prompt.
For the model-only setting, context-limit failures are reported separately
and are not counted as incomplete.
}
\label{tab:evaluation_settings}
\end{table*}

\clearpage
\section{Detailed PVSR Change from Static Tool-Augmented Agent to Internet-Enabled Open-World Agent}
\label{app:score_change}

\begin{table}[!htbp]
\centering
\caption{
Task-level success transitions from the Static Tool-Augmented setting
to the Internet-Enabled Open-World setting.
Success$\rightarrow$Fail counts tasks that lose success, while
Fail$\rightarrow$Success counts tasks that newly become successful.
}
\label{tab:success_transitions}
\small
\begin{tabular}{lrrrrr}
\toprule
\textbf{Agent}
& \textbf{Static Success}
& \textbf{Success$\rightarrow$Fail}
& \textbf{Fail$\rightarrow$Success}
& \textbf{Open-World Success}
& \textbf{$\Delta$} \\
\midrule
Claude Opus 5
& 81 & 7  & 6  & 80 & $-1$  \\

GPT-5.6 Sol
& 72 & 17 & 7  & 62 & $-10$ \\

DeepSeek-V4.1 Flash
& 46 & 10 & 21 & 57 & $+11$ \\

GPT-5.6 Luna
& 23 & 11 & 15 & 27 & $+4$  \\

GLM-5.3 Flash
& 25 & 9  & 6  & 22 & $-3$  \\
\midrule
\textbf{All 5}
& \textbf{247}
& \textbf{54}
& \textbf{55}
& \textbf{248}
& \textbf{$+1$} \\
\bottomrule
\end{tabular}
\end{table}



\lstset{
  basicstyle=\ttfamily\scriptsize,
  columns=fullflexible,
  keepspaces=true,
  frame=single,
  breaklines=true,
  xleftmargin=0.5em,
  xrightmargin=0.5em,
  aboveskip=0.5em,
  belowskip=0.5em
}


\clearpage
\section{Case Studies of Common Reasoning Failures}
\label{app:reasoning-failure-cases}

This appendix presents three representative case studies of recurring reasoning failures in indirect-call resolution, corresponding to the three common failure modes discussed in \autoref{sec:failure}.

\begin{table}[!htbp]
    \centering
    \caption{Overview of the three case studies.}
    \label{tab:case-overview}
    \small
    \begin{tabularx}{\linewidth}{@{}l X l l@{}}
        \toprule
        Case & Reasoning challenge & Call site & Task \\
        \midrule

        C1 &
        Recovering the value of a mutable global
        function-pointer slot &
        \texttt{0x17103} &
        \texttt{libxml2\_gcc\_O2} \\

        C2 &
        Mapping packed function pointers to the correct structure field &
        \texttt{0xaa504} &
        \texttt{libxml2\_gcc\_O2} \\

        C3 &
        Recovering an interprocedural default target and preserving it
        through branch reconvergence &
        \texttt{0x714d9} &
        \texttt{njs\_gcc\_O2} \\

        \bottomrule
    \end{tabularx}
\end{table}


\clearpage

\subsection{C1: Recovering the Provenance of a Global
Function-Pointer Slot}
\label{app:case-c1}
\suppressfloats[t]

\PP{Call site and ground truth.}
The first case occurs at call site \texttt{0x17103} in
\texttt{libxml2\_gcc\_O2}. The indirect call reads its target from the
global function-pointer slot at \texttt{0x1380e0}:

\begin{figure}[!htbp]
    \centering

    \begin{subfigure}[t]{0.48\linewidth}
    \caption{Indirect call site.}
\begin{lstlisting}
170ff: mov    rbp,rsi
17102: push   rbx
17103: call   QWORD PTR [rip+0x120fd7]
                 # 1380e0
17109: test   rax,rax
\end{lstlisting}
    \end{subfigure}
    \hfill
    \begin{subfigure}[t]{0.48\linewidth}
    \caption{Setter that overwrites the slot.}
\begin{lstlisting}
76740: test   rdi,rdi
76743: je     76770
76745: mov    QWORD PTR [rip+0xc199c],rdi
                 # 1380e8
7674e: mov    QWORD PTR [rip+0xc198b],rsi
                 # 1380e0
76755: mov    QWORD PTR [rip+0xc197c],rsi
                 # 1380d8
\end{lstlisting}
    \end{subfigure}

    \vspace{0.5em}

    \begin{subfigure}[t]{0.48\linewidth}
    \caption{Caller supplying the replacement pointer.}
\begin{lstlisting}
d71b2: lea    rdx,[rip+0xffffffffffffff17]
             # d70d0
d71b9: lea    rsi,[rip+0xffffffffffffff70]
             # d7130
d71c0: jmp    76720
\end{lstlisting}
    \end{subfigure}
    \hfill
    \begin{subfigure}[t]{0.48\linewidth}
    \caption{In-binary allocation wrapper.}
\begin{lstlisting}
d7134: mov    rdx,QWORD PTR [rip+0x6254d]
             # 139688
d7140: mov    rax,QWORD PTR [rip+0x62549]
             # 139690
d7147: add    rax,0x1
d714b: mov    QWORD PTR [rip+0x6253e],rax
d7157: sub    rsp,0x8
d715b: call   167d0 <malloc@plt>
\end{lstlisting}
    \end{subfigure}

    \caption{Case C1. The indirect call reads a global slot whose
    initial relocation references \texttt{malloc}, but the slot is
    subsequently overwritten with the address of an in-binary
    allocation wrapper.}
    \label{fig:c1-code}
\end{figure}

The ground-truth target is \texttt{0xd7130}, a function entry inside
the binary.

The source of ambiguity is the load-time relocation associated with
the global slot:

\begin{lstlisting}
00000000001380e0 R_X86_64_64 malloc@GLIBC_2.2.5
\end{lstlisting}

Taken in isolation, this relocation makes the indirect call appear to
leave the binary and enter libc. However, the relocation specifies
only the \emph{initial} value of the slot. The program later modifies
the same location:

\begin{lstlisting}
76740: test   rdi,rdi
76743: je     76770
76745: mov    QWORD PTR [rip+0xc199c],rdi
             # 1380e8
7674e: mov    QWORD PTR [rip+0xc198b],rsi
             # 1380e0
76755: mov    QWORD PTR [rip+0xc197c],rsi
             # 1380d8
\end{lstlisting}

Tracing the value in \texttt{rsi} back to the caller identifies the
replacement function pointer:

\begin{lstlisting}
d71b2: lea    rdx,[rip+0xffffffffffffff17]
             # d70d0
d71b9: lea    rsi,[rip+0xffffffffffffff70]
             # d7130
d71c0: jmp    76720
\end{lstlisting}

Thus, \texttt{0xd7130} is written into the slot later read at
\texttt{0x17103}. The target is an in-binary allocation wrapper that
updates the harness allocation state before invoking
\texttt{malloc@plt}:

\begin{lstlisting}
d7134: mov    rdx,QWORD PTR [rip+0x6254d]
             # 139688
d7140: mov    rax,QWORD PTR [rip+0x62549]
             # 139690
d7147: add    rax,0x1
d714b: mov    QWORD PTR [rip+0x6253e],rax
d7157: sub    rsp,0x8
d715b: call   167d0 <malloc@plt>
\end{lstlisting}

The complete static provenance chain is therefore

\[
\texttt{0xd71b9}
\rightarrow
\texttt{0x7674e}
\rightarrow
\texttt{[0x1380e0]}
\rightarrow
\texttt{0x17103}.
\]

\PP{How the agents reasoned.}
Failing analyses interpret the relocation as evidence that the
indirect call ultimately targets an external libc function. One
analysis therefore returns no in-binary target, while others identify
the corresponding \texttt{malloc@plt} entry.

This reasoning is locally plausible because the relocation genuinely
associates \texttt{0x1380e0} with \texttt{malloc}, and the PLT entry
is the binary's direct representation of that external function. The
failure arises because the analysis treats the relocation as the
pointer's final provenance rather than as its initialization state.

The missing step is a store-side analysis. After identifying
\texttt{0x1380e0} as the slot used by the indirect call, the analysis
must determine whether instructions in \texttt{.text} subsequently
write to that location. The store at \texttt{0x7674e} and the
\texttt{lea} at \texttt{0xd71b9} then expose the actual in-binary
target.

\begin{figure}[!htbp]
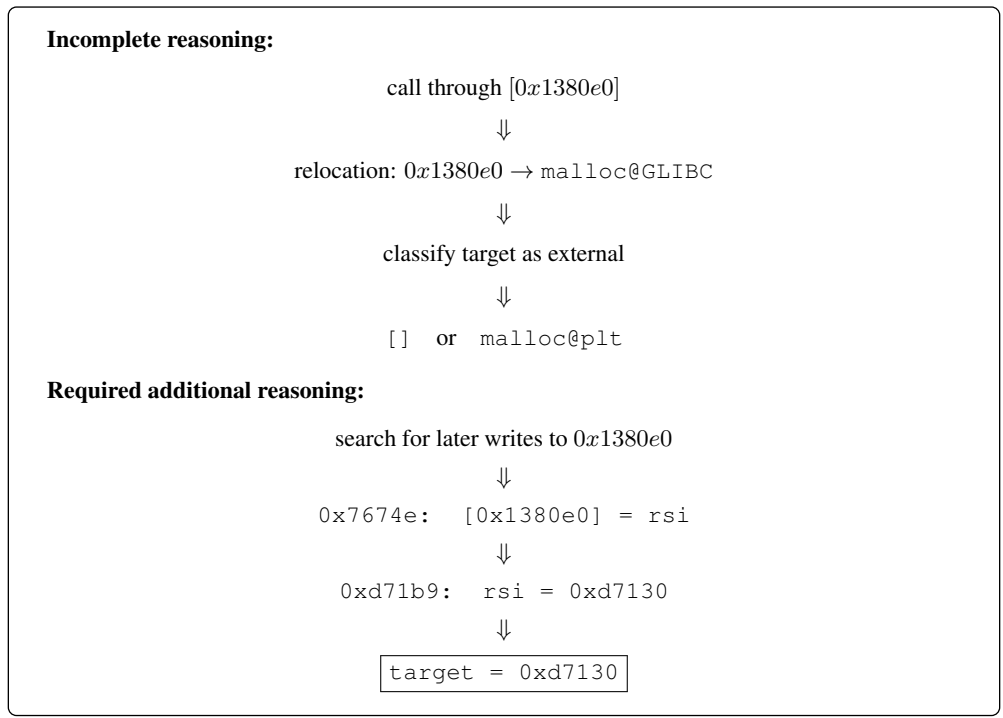

    \centering

    \begin{tcolorbox}[
        width=0.94\linewidth,
        colback=white,
        colframe=black,
        boxrule=0.5pt
    ]
    \small

    \textbf{Incomplete reasoning:}

    \[
    \begin{array}{c}
    \text{call through }[0x1380e0]
    \\[2mm]
    \Downarrow
    \\[2mm]
    \text{relocation: }
    0x1380e0 \rightarrow \texttt{malloc@GLIBC}
    \\[2mm]
    \Downarrow
    \\[2mm]
    \text{classify target as external}
    \\[2mm]
    \Downarrow
    \\[2mm]
    \texttt{[]} \quad\text{or}\quad \texttt{malloc@plt}
    \end{array}
    \]

    \medskip

    \textbf{Required additional reasoning:}

    \[
    \begin{array}{c}
    \text{search for later writes to }0x1380e0
    \\[1.5mm]
    \Downarrow
    \\[1.5mm]
    \texttt{0x7674e: [0x1380e0] = rsi}
    \\[1.5mm]
    \Downarrow
    \\[1.5mm]
    \texttt{0xd71b9: rsi = 0xd7130}
    \\[1.5mm]
    \Downarrow
    \\[1.5mm]
    \boxed{\texttt{target = 0xd7130}}
    \end{array}
    \]

    \end{tcolorbox}

    \caption{Reasoning failure in Case C1. The incomplete analysis
    terminates at relocation metadata, whereas the complete analysis
    additionally tracks subsequent writes to the mutable slot.}
    \label{fig:c1-trajectory}
\end{figure}

\PP{Why the reasoning fails.}
C1 is therefore a provenance error rather than a failure to locate the
relevant memory operand. The call site clearly identifies the slot at
\texttt{0x1380e0}; the error occurs because the analysis assigns too
much authority to the slot's load-time relocation.

The critical distinction is between \emph{where a pointer is
initialized} and \emph{which value reaches the call site}. For mutable
global state, these are not necessarily the same. In this case, a
short static chain resolves the target, but omitting the store-side
analysis causes the initial external value to be mistaken for the
effective indirect-call target.


\clearpage

\subsection{C2: Recovering the Correct Structure Field from a
Vectorized Initialization}
\label{app:case-c2}
\suppressfloats[t]

\PP{Call site and ground truth.}
The second case occurs at \texttt{0xaa504} in
\texttt{libxml2\_gcc\_O2}. The indirect target is loaded from offset
\texttt{+0xf0}:

\begin{figure}[!htbp]
    \centering

    \begin{subfigure}[t]{0.48\linewidth}
    \caption{Indirect call through field \texttt{+0xf0}.}
\begin{lstlisting}
aa4d3: mov    rax,QWORD PTR [rbx]
aa4d6: test   rax,rax
aa4db: mov    rax,QWORD PTR [rax+0xf0]
aa4e2: test   rax,rax
...
aa4f9: mov    rsi,QWORD PTR [rbx+0x120]
aa500: mov    rdi,QWORD PTR [rbx+0x8]
aa504: call   rax
\end{lstlisting}
    \end{subfigure}
    \hfill
    \begin{subfigure}[t]{0.48\linewidth}
    \caption{Vectorized structure initialization.}
\begin{lstlisting}
d2460: lea    rdx,[rip+0xffffffffffffca69]
             # ceed0
d2467: lea    rdi,[rip+0xfffffffffffff612]
             # d1a80
d246e: mov    QWORD PTR [rax+0xf8],0x0
d2479: movq   xmm7,rdx
d247e: movq   xmm3,rdi
d2488: punpcklqdq xmm3,xmm7
d248c: movups XMMWORD PTR [rax+0xe8],xmm3
d2493: jmp    d22f6
\end{lstlisting}
    \end{subfigure}

    \caption{Case C2. Two adjacent function-pointer fields are
    initialized simultaneously by a 16-byte SSE store.}
    \label{fig:c2-code}
\end{figure}

The ground-truth target is \texttt{0xceed0}. The two candidate
functions are materialized next to one another:

\begin{lstlisting}
d2460: lea    rdx,[rip+0xffffffffffffca69]
             # ceed0
d2467: lea    rdi,[rip+0xfffffffffffff612]
             # d1a80
\end{lstlisting}

They are then copied into SIMD registers and packed:

\begin{lstlisting}
d2479: movq   xmm7,rdx
d247e: movq   xmm3,rdi
d2488: punpcklqdq xmm3,xmm7
\end{lstlisting}

The resulting register layout is

\[
\texttt{xmm3}
=
[
\texttt{0xd1a80}
\mid
\texttt{0xceed0}
].
\]

The following 16-byte store begins at offset \texttt{+0xe8}:

\begin{lstlisting}
d248c: movups XMMWORD PTR [rax+0xe8],xmm3
\end{lstlisting}

Therefore,

\[
\begin{aligned}
\texttt{object+0xe8} &\leftarrow \texttt{0xd1a80},\\
\texttt{object+0xf0} &\leftarrow \texttt{0xceed0}.
\end{aligned}
\]

Since the indirect call reads from \texttt{+0xf0}, the target is

\[
\boxed{\texttt{0xceed0}}.
\]

\PP{How the agents reasoned.}
Several analyses recognize the correct handler initialization but do
not preserve the exact association between the two function pointers
and their destination fields. In particular, more than one analysis
selects \texttt{0xd1a80}, which is not an unrelated function: it is
the handler stored in the immediately adjacent field at
\texttt{+0xe8}.

This makes the error especially informative. The analysis has already
recovered the relevant handler pair,

\[
\{
\texttt{0xd1a80},
\texttt{0xceed0}
\},
\]

but the two values become indistinguishable unless the SIMD operation
is interpreted precisely. Looking only at the nearby \texttt{lea}
instructions reveals both candidates but does not determine which one
reaches \texttt{+0xf0}.

The decisive information is the operand order of
\texttt{punpcklqdq} together with the destination address of the
subsequent \texttt{movups}. These instructions establish the exact
mapping from each function pointer to its 8-byte structure field.

\begin{figure}[!htbp]
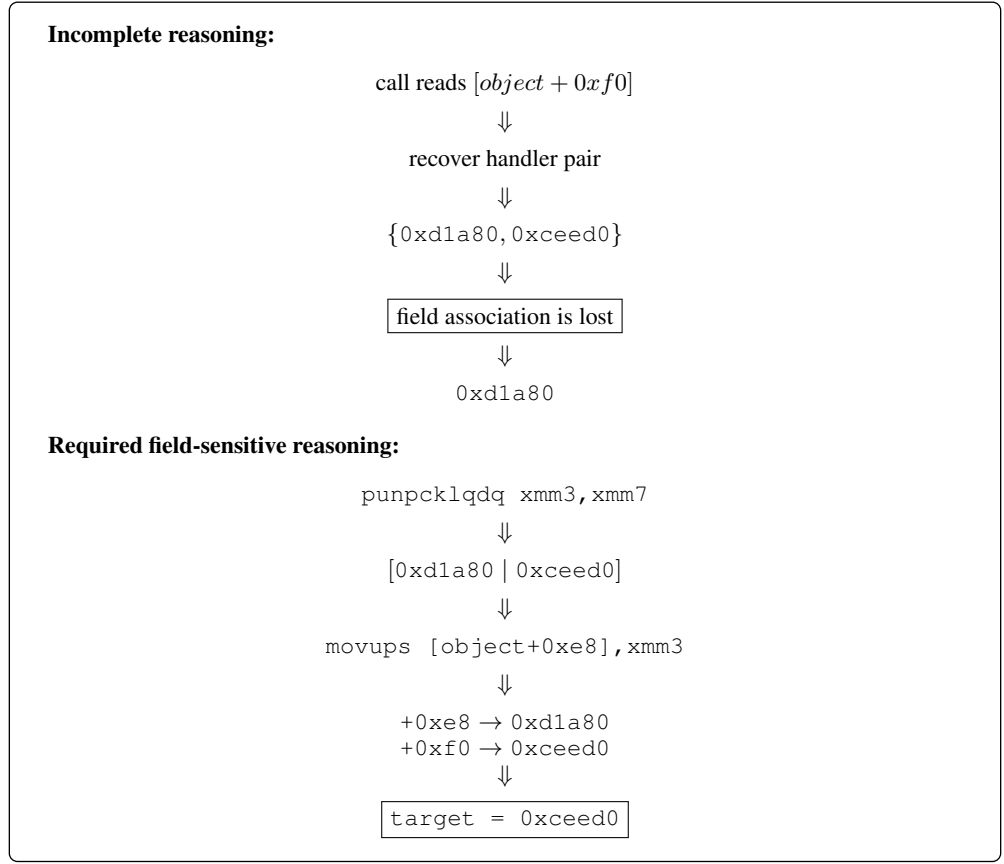

    \centering

    \begin{tcolorbox}[
        width=0.94\linewidth,
        colback=white,
        colframe=black,
        boxrule=0.5pt
    ]
    \small

    \textbf{Incomplete reasoning:}

    \[
    \begin{array}{c}
    \text{call reads }[object+0xf0]
    \\[1.5mm]
    \Downarrow
    \\[1.5mm]
    \text{recover handler pair}
    \\[1.5mm]
    \Downarrow
    \\[1.5mm]
    \{
    \texttt{0xd1a80},
    \texttt{0xceed0}
    \}
    \\[1.5mm]
    \Downarrow
    \\[1.5mm]
    \boxed{\text{field association is lost}}
    \\[1.5mm]
    \Downarrow
    \\[1.5mm]
    \texttt{0xd1a80}
    \end{array}
    \]

    \medskip

    \textbf{Required field-sensitive reasoning:}

    \[
    \begin{array}{c}
    \texttt{punpcklqdq xmm3,xmm7}
    \\[1.5mm]
    \Downarrow
    \\[1.5mm]
    [
    \texttt{0xd1a80}
    \mid
    \texttt{0xceed0}
    ]
    \\[1.5mm]
    \Downarrow
    \\[1.5mm]
    \texttt{movups [object+0xe8],xmm3}
    \\[1.5mm]
    \Downarrow
    \\[1.5mm]
    \begin{array}{l}
    \texttt{+0xe8} \rightarrow \texttt{0xd1a80}\\
    \texttt{+0xf0} \rightarrow \texttt{0xceed0}
    \end{array}
    \\[1.5mm]
    \Downarrow
    \\[1.5mm]
    \boxed{\texttt{target = 0xceed0}}
    \end{array}
    \]

    \end{tcolorbox}

    \caption{Reasoning failure in Case C2. The handler family is
    recovered, but correct resolution additionally requires
    reconstructing the exact byte-level mapping of the vectorized
    store.}
    \label{fig:c2-trajectory}
\end{figure}

\PP{Why the reasoning fails.}
C2 is a field-sensitivity failure caused by compiler vectorization.
A scalar analysis searching only for stores of the form

\begin{lstlisting}
mov QWORD PTR [reg+offset], function_pointer
\end{lstlisting}

does not recover either assignment because both fields are written by
a single \texttt{movups} operation.

The incorrect target remains structurally plausible: it belongs to the
same handler pair, is materialized in the same initialization
sequence, and is written by the same vector store as the correct
target. The only incorrect step is the mapping between the packed
pointer value and the exact structure offset.

Thus, recognizing the correct callback family is not sufficient.
Indirect-call resolution must preserve field-level correspondence
through compiler transformations such as SIMD packing and vectorized
stores.


\clearpage

\subsection{C3: Interprocedural Pointer Provenance and
Path Reconvergence}
\label{app:case-c3}
\suppressfloats[t]

\PP{Call site and ground truth.}
The third case occurs at \texttt{0x714d9} in
\texttt{njs\_gcc\_O2}. The ground-truth target set contains four
in-binary function entries:

\[
T =
\{
\texttt{0x53150},
\texttt{0x532c0},
\texttt{0x53430},
\texttt{0x53440}
\}.
\]

The indirect function pointer is loaded from offset \texttt{+0x18} of
an object supplied as the fourth argument:

\begin{figure}[!htbp]
    \centering

    \begin{subfigure}[t]{0.48\linewidth}
    \caption{Indirect call through argument field \texttt{+0x18}.}
\begin{lstlisting}
71470: endbr64
71474: push   r13
71476: mov    r13,rcx
        ; r13 = fourth argument
...
714b0: mov    rcx,QWORD PTR [r13+0x18]
714b4: cmp    rcx,QWORD PTR [rip+0x4faed]
             # c0fa8
714ce: je     71500
714d0: mov    rdx,rsp
714d3: mov    rsi,r12
714d6: mov    rdi,rbp
714d9: call   rcx
\end{lstlisting}
    \end{subfigure}
    \hfill
    \begin{subfigure}[t]{0.48\linewidth}
    \caption{Control-flow reconvergence.}
\begin{lstlisting}
71500: xor    eax,eax
71502: cmp    QWORD PTR [rsp+0x18],0x0
71508: je     714d0
        ; rejoins call path
        ; rcx remains unchanged

7150a: jmp    714db
        ; only this path skips call
\end{lstlisting}
    \end{subfigure}

    \caption{Case C3. The function pointer is obtained from an object
    field supplied by the caller. The equality branch appears to
    exclude one candidate, but one successor later reconverges with
    the indirect-call path.}
    \label{fig:c3-callsite}
\end{figure}

The complete provenance of the field is not contained within this
function. The object arrives as a parameter, so a purely
intraprocedural backward slice reaches the incoming argument without
establishing the complete set of values that may occupy
\texttt{+0x18}.

\PP{Recovering the statically initialized default.}
One target, \texttt{0x53150}, is particularly difficult because it is
not introduced by an ordinary scalar store to offset \texttt{+0x18}.
Instead, related functions compare the same object field against this
address:

\begin{figure}[!htbp]
    \centering

    \begin{subfigure}[t]{0.48\linewidth}
    \caption{Related function at \texttt{0x711f0}.}
\begin{lstlisting}
71289: lea    rax,[rip+0xfffffffffffe1ec0]
             # 53150
7129a: cmp    QWORD PTR [r13+0x18],rax
7129e: je     71430
\end{lstlisting}
    \end{subfigure}
    \hfill
    \begin{subfigure}[t]{0.48\linewidth}
    \caption{Related function at \texttt{0x71a10}.}
\begin{lstlisting}
71aa6: lea    rcx,[rip+0xfffffffffffe16a3]
             # 53150
71ab1: cmp    QWORD PTR [r14+0x18],rcx
71ab5: je     71b60
\end{lstlisting}
    \end{subfigure}

    \caption{Related functions expose \texttt{0x53150} as the default
    value associated with field \texttt{+0x18}.}
    \label{fig:c3-siblings}
\end{figure}

The same value also appears in the relocation for the global used by
the local comparison:

\begin{lstlisting}
00000000000c0fa8 R_X86_64_RELATIVE *ABS*+0x0000000000053150
\end{lstlisting}

Therefore,

\[
\texttt{[0xc0fa8]} = \texttt{0x53150}.
\]

This establishes \texttt{0x53150} as a statically initialized default
for the relevant field. A store-enumeration analysis restricted to

\begin{lstlisting}
mov QWORD PTR [reg+0x18], function_pointer
\end{lstlisting}

does not recover this value. The analysis must instead combine
evidence from related functions with relocation metadata.

\PP{How the agents reasoned.}
Two characteristic reasoning patterns appear at this site.

Some agents stop after tracing the function pointer to
\texttt{[r13+0x18]}. Because \texttt{r13} originates from an incoming
parameter, the local function provides no concrete producer, and the
analysis terminates without resolving a target set. This is a typical
failure of purely intraprocedural provenance reasoning.

Most of the stronger analyses proceed further and recover three of the
four ground-truth targets. However, they omit the same fourth target,
\texttt{0x53150}. The omitted value is precisely the statically
initialized default identified by the relocation and the related
functions.

A plausible reason for this omission is the local comparison immediately
before the indirect call:

\begin{lstlisting}
714b4: cmp    rcx,QWORD PTR [rip+0x4faed]
             # c0fa8
714ce: je     71500
\end{lstlisting}

Because \texttt{[0xc0fa8]} is initialized to \texttt{0x53150}, a local
reading suggests

\[
\texttt{rcx} = \texttt{0x53150}
\quad\Longrightarrow\quad
\text{execution branches away from the call}.
\]

Under this interpretation, \texttt{0x53150} appears to be a possible
field value but not a feasible target at \texttt{0x714d9}.

The flaw is that the branch destination itself can return to the call
path:

\begin{lstlisting}
71500: xor    eax,eax
71502: cmp    QWORD PTR [rsp+0x18],0x0
71508: je     714d0
\end{lstlisting}

The jump at \texttt{0x71508} reaches \texttt{0x714d0} without
overwriting \texttt{rcx}. Therefore,

\[
\texttt{0x714ce}
\rightarrow
\texttt{0x71500}
\rightarrow
\texttt{0x71508}
\rightarrow
\texttt{0x714d0}
\rightarrow
\texttt{0x714d9}
\]

is a valid path reaching the indirect call with

\[
\texttt{rcx}=\texttt{0x53150}.
\]

Thus, the candidate is valid even though the first branch initially
appears to exclude it.

\begin{figure}[!htbp]
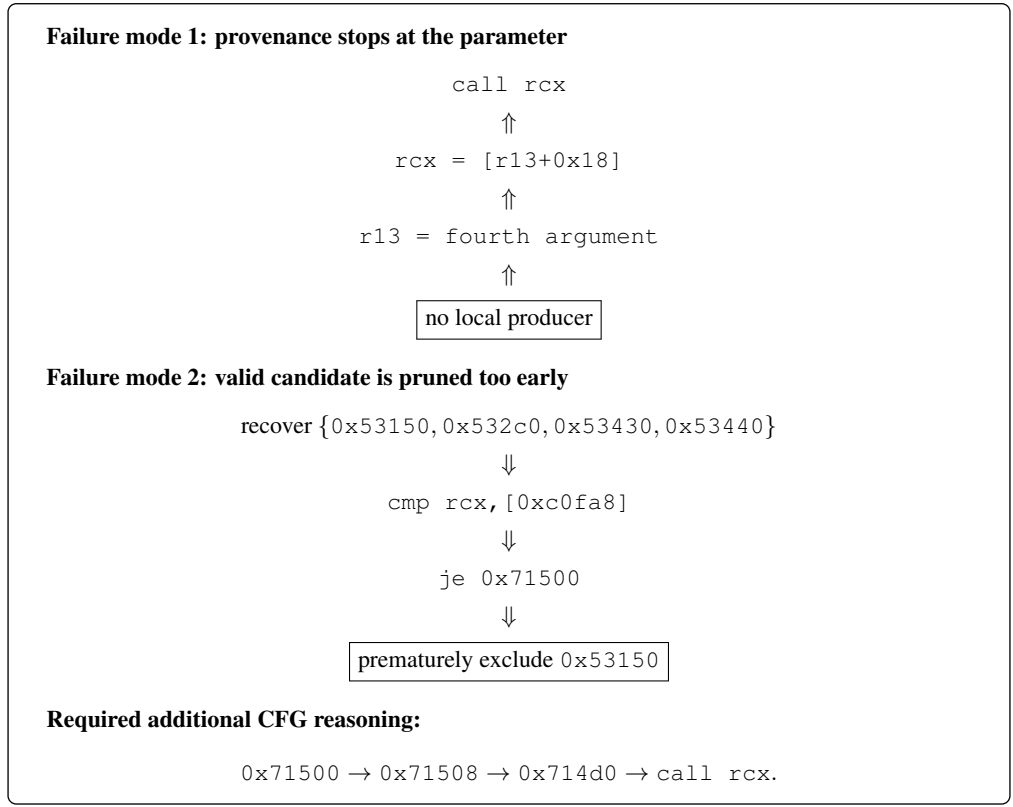

    \centering

    \begin{tcolorbox}[
        width=0.95\linewidth,
        colback=white,
        colframe=black,
        boxrule=0.5pt
    ]
    \small

    \textbf{Failure mode 1: provenance stops at the parameter}

    \[
    \begin{array}{c}
    \texttt{call rcx}
    \\[1.5mm]
    \Uparrow
    \\[1.5mm]
    \texttt{rcx = [r13+0x18]}
    \\[1.5mm]
    \Uparrow
    \\[1.5mm]
    \texttt{r13 = fourth argument}
    \\[1.5mm]
    \Uparrow
    \\[1.5mm]
    \boxed{\text{no local producer}}
    \end{array}
    \]

    \medskip

    \textbf{Failure mode 2: valid candidate is pruned too early}

    \[
    \begin{array}{c}
    \text{recover }
    \{
    \texttt{0x53150},
    \texttt{0x532c0},
    \texttt{0x53430},
    \texttt{0x53440}
    \}
    \\[1.5mm]
    \Downarrow
    \\[1.5mm]
    \texttt{cmp rcx,[0xc0fa8]}
    \\[1.5mm]
    \Downarrow
    \\[1.5mm]
    \texttt{je 0x71500}
    \\[1.5mm]
    \Downarrow
    \\[1.5mm]
    \boxed{\text{prematurely exclude }\texttt{0x53150}}
    \end{array}
    \]

    \medskip

    \textbf{Required additional CFG reasoning:}

    \[
    \texttt{0x71500}
    \rightarrow
    \texttt{0x71508}
    \rightarrow
    \texttt{0x714d0}
    \rightarrow
    \texttt{call rcx}.
    \]

    \end{tcolorbox}

    \caption{Two reasoning failures in Case C3. A shallow
    intraprocedural analysis can stop at the incoming parameter, while
    a stronger analysis can recover the default candidate but remove
    it using an incomplete interpretation of the local CFG.}
    \label{fig:c3-trajectory}
\end{figure}

\PP{Why the reasoning fails.}
C3 requires three forms of reasoning simultaneously.

First, the function pointer originates outside the local function, so
the analysis must cross a function boundary rather than terminate when
the backward slice reaches an argument.

Second, \texttt{0x53150} is a statically initialized default rather
than a value exposed by an ordinary scalar store. Recovering it
requires correlating the same object field across related functions
and incorporating relocation metadata.

Third, candidate discovery must be separated from path-feasibility
filtering. Even after identifying \texttt{0x53150}, a shallow
interpretation of the first equality branch makes the target appear
unreachable. Following the branch target further reveals that the
control flow reconverges with the call path while preserving the
candidate in \texttt{rcx}.

The case therefore illustrates two distinct ways reasoning can fail:
the analysis can stop too early while tracing provenance, or it can
recover the correct candidate and then eliminate it using an
incomplete control-flow argument.

\clearpage
\section{Case Studies of Unique Pass Cases}
\label{app:successful-reasoning-strategies}

This appendix presents two representative case studies illustrating
the complementary reasoning strategies discussed in \autoref{sec:failure}.
Specifically, Claude Opus 5 explicitly models the relation between
function-pointer assignments, structure fields, and indirect calls;
and GPT-5.6 Sol identifies the relevant callback relation and resolves
its members exhaustively before committing.

\begin{table}[!htbp]
    \centering
    \caption{Overview of the two representative reasoning strategies.}
    \label{tab:strategy-overview}
    \small
    \begin{tabularx}{\linewidth}{@{}l l X l l@{}}
        \toprule
        Case & Model & Reasoning strategy & Call site & Task \\
        \midrule

        S1 &
        Claude Opus 5 &
        Model function-pointer assignment, structure field, and call
        site as one relation &
        \texttt{0xb616} &
        \texttt{openjpeg\_gcc\_O2} \\

        S2 &
        GPT-5.6 Sol &
        Identify the callback relation and resolve its members
        exhaustively before committing &
        \texttt{0x58be5} &
        \texttt{wabt\_gcc\_O0} \\

        \bottomrule
    \end{tabularx}
\end{table}


\clearpage

\subsection{S1: Claude Opus 5---Field-Indexed Modeling of
Function-Pointer Assignments}
\label{app:strategy-s1}
\suppressfloats[t]

\PP{Call site and target relation.}
The first case comes from \texttt{openjpeg\_gcc\_O2}. At
\texttt{0xb616}, the indirect call reads a function pointer from
structure offset \texttt{+0x30}:

\begin{figure}[!htbp]
    \centering

    \begin{subfigure}[t]{0.46\linewidth}
    \caption{Indirect call through field \texttt{+0x30}.}
\begin{lstlisting}
b616: call QWORD PTR [rbx+0x30]

Observed Ground-truth targets:
    0x306e0
    0x4cef0
    0x4cf30
\end{lstlisting}
    \end{subfigure}
    \hfill
    \begin{subfigure}[t]{0.51\linewidth}
    \caption{Field-indexed callback sets.}
\begin{lstlisting}
STREAM_FREE =
  [0x109f0, 0x4c6f0, 0x4c730]       # +0x08

STREAM_READ =
  [0xb650, 0x10a80, 0x30700,
   0x4c470, 0x4ce10]                 # +0x18

STREAM_WRITE =
  [0xb680, 0x10a60,
   0x4c5f0, 0x4c660]                 # +0x20

STREAM_SKIP =
  [0xb660, 0x10a30, 0x306f0,
   0x4c530, 0x4c580]                 # +0x28

STREAM_SEEK =
  [0xb670, 0x10a00, 0x306e0,
   0x4cef0, 0x4cf30]                 # +0x30
\end{lstlisting}
    \end{subfigure}

    \caption{Case S1. Claude Opus 5 indexes function-pointer
    assignments by structure offset, allowing a call through
    \texttt{+0x30} to be associated specifically with the
    \texttt{STREAM\_SEEK} callback family.}
    \label{fig:s1-field-index}
\end{figure}

The important feature of Claude Opus 5's analysis is that it does not
represent the stream callbacks as one undifferentiated collection of
function pointers. Instead, a whole-binary scan records which function
addresses are assigned to each structure offset. The relevant relation
is therefore

\[
\text{function-pointer assignment}
\rightarrow
\text{structure field}
\rightarrow
\text{indirect call site}.
\]

For the call at \texttt{0xb616}, the load from
\texttt{[rbx+0x30]} selects the callback set associated specifically
with offset \texttt{+0x30}. This set contains all three observed
ground-truth targets.

\PP{How Claude Opus 5 reasons.}
The key step is to use the structure-field offset as the join key
between stores and indirect calls. A store discovered in one function
can therefore be associated with a call in another function as long as
both refer to the same field.

\begin{figure}[!htbp]
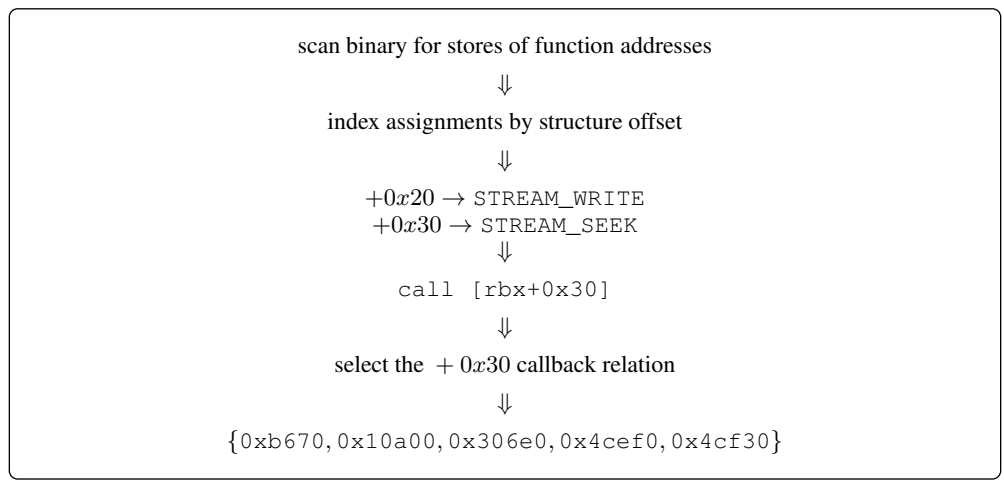

    \centering

    \begin{tcolorbox}[
        width=0.94\linewidth,
        colback=white,
        colframe=black,
        boxrule=0.5pt
    ]
    \small

    \[
    \begin{array}{c}
    \text{scan binary for stores of function addresses}
    \\[1.5mm]
    \Downarrow
    \\[1.5mm]
    \text{index assignments by structure offset}
    \\[1.5mm]
    \Downarrow
    \\[1.5mm]
    \begin{array}{c}
    +0x20 \rightarrow \texttt{STREAM\_WRITE}\\
    +0x30 \rightarrow \texttt{STREAM\_SEEK}
    \end{array}
    \\[1.5mm]
    \Downarrow
    \\[1.5mm]
    \texttt{call [rbx+0x30]}
    \\[1.5mm]
    \Downarrow
    \\[1.5mm]
    \text{select the }+0x30\text{ callback relation}
    \\[1.5mm]
    \Downarrow
    \\[1.5mm]
    \{
      \texttt{0xb670},
      \texttt{0x10a00},
      \texttt{0x306e0},
      \texttt{0x4cef0},
      \texttt{0x4cf30}
    \}
    \end{array}
    \]

    \end{tcolorbox}

    \caption{Claude Opus 5's reasoning strategy in S1. The
    structure-field offset provides a stable relation between
    assignments and uses even when they occur in different functions.}
    \label{fig:s1-trajectory}
\end{figure}

\PP{Why the strategy succeeds.}
Claude Opus 5's field-indexed representation preserves distinctions
between adjacent members of the same callback structure. For example,
the \texttt{STREAM\_WRITE} callbacks at \texttt{+0x20} and the
\texttt{STREAM\_SEEK} callbacks at \texttt{+0x30} belong to the same
overall object but remain separate relations.

More importantly, this representation naturally crosses function
boundaries. The assignment and the eventual indirect call need not
occur in the same function: once stores are indexed by field offset,
every call that reads that field can reuse the recovered assignment
set. This allows Claude Opus 5 to connect function-pointer
initialization with distant indirect uses while retaining field-level
precision.


\clearpage

\subsection{S2: GPT-5.6 Sol---Exhaustive Resolution of a
Callback Relation}
\label{app:strategy-s2}
\suppressfloats[t]

\PP{Call site and target relation.}
The second case occurs at \texttt{0x58be5} in
\texttt{wabt\_gcc\_O0}. The site is a C++ virtual call through
vtable slot \texttt{+0x8}:

\begin{figure}[!htbp]
    \centering

    \begin{subfigure}[t]{0.50\linewidth}
    \caption{Virtual call site.}
\begin{lstlisting}
58bc3: mov    rax,QWORD PTR [rbp-0x10]
58bc7: mov    rax,QWORD PTR [rax+0x8]
58bcf: mov    rax,QWORD PTR [rbp-0x10]
58bd3: test   rax,rax
58bd8: mov    rdx,QWORD PTR [rax]
        ; load vptr
58bdb: add    rdx,0x8
        ; vtable slot +0x8
58bdf: mov    rdx,QWORD PTR [rdx]
58be2: mov    rdi,rax
58be5: call   rdx
\end{lstlisting}
    \end{subfigure}
    \hfill
    \begin{subfigure}[t]{0.46\linewidth}
    \caption{Ground-truth override set.}
\begin{lstlisting}
0x1109f6
0x1118de
0x112154
0x112210
0x1122cc
0x112388
0x112444
0x1124ea
0x11c4b4
0x11cbe2
0x11ce28
\end{lstlisting}
    \end{subfigure}

    \caption{Case S2. GPT-5.6 Sol identifies the fixed vtable slot
    \texttt{+0x8} and resolves the complete set of relevant overrides
    before committing to the target set.}
    \label{fig:s2-vtable}
\end{figure}

The target relation is more specific than simply identifying functions
that occur in vtables. The call identifies a particular virtual slot,
and the analysis must determine which concrete classes contribute an
implementation at that same slot.

GPT-5.6 Sol first reconstructs the surrounding relations---including
RTTI objects, primary vtables, delegate families, and callback
propagation through callers---and then resolves the members belonging
to the relevant relation. At \texttt{0x58be5}, this yields the
complete set of eleven overrides of slot \texttt{+0x8}.

\PP{How GPT-5.6 Sol reasons.}
The defining behavior is that GPT-5.6 Sol closes the callback relation
before committing to an answer. It first identifies which relation the
call belongs to and then enumerates the members of that relation
exhaustively.

\begin{figure}[!htbp]
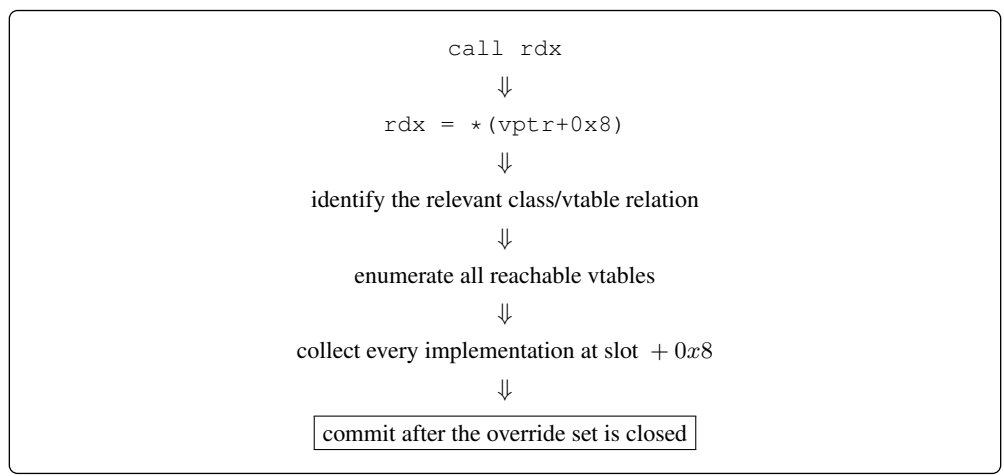

    \centering

    \begin{tcolorbox}[
        width=0.94\linewidth,
        colback=white,
        colframe=black,
        boxrule=0.5pt
    ]
    \small

    \[
    \begin{array}{c}
    \texttt{call rdx}
    \\[1.5mm]
    \Downarrow
    \\[1.5mm]
    \texttt{rdx = *(vptr+0x8)}
    \\[1.5mm]
    \Downarrow
    \\[1.5mm]
    \text{identify the relevant class/vtable relation}
    \\[1.5mm]
    \Downarrow
    \\[1.5mm]
    \text{enumerate all reachable vtables}
    \\[1.5mm]
    \Downarrow
    \\[1.5mm]
    \text{collect every implementation at slot }+0x8
    \\[1.5mm]
    \Downarrow
    \\[1.5mm]
    \boxed{\text{commit after the override set is closed}}
    \end{array}
    \]

    \end{tcolorbox}

    \caption{GPT-5.6 Sol's reasoning strategy in S2. Candidate
    generation is organized around a concrete callback relation whose
    members are resolved exhaustively before an answer is emitted.}
    \label{fig:s2-trajectory}
\end{figure}

\PP{Why the strategy succeeds.}
GPT-5.6 Sol's ordering avoids two complementary errors. Committing
before the relation is fully resolved can omit valid overrides, while
collecting loosely related functions can introduce targets that do not
belong to the relevant virtual slot.

By first identifying the relation and then closing its membership,
GPT-5.6 Sol can be both complete and selective. At
\texttt{0x58be5}, this strategy recovers all eleven targets. More
generally, it produces exact target sets at all eight closed-set sites
in this task without adding extra targets.

The case therefore shows that exhaustive resolution does not
necessarily require broad prediction. GPT-5.6 Sol applies
exhaustiveness within a well-defined callback relation, allowing
completeness and precision to reinforce each other.

\clearpage

\section{Agent Harness Ablation}
\label{app:agent_scaffold}
\begin{table}[!htbp]
    \centering
    \small
    \setlength{\tabcolsep}{6pt}
    \caption{Effect of agent harness on PVSR.}
    \label{tab:rq2_scaffold}
    \begin{tabular}{llcc}
        \toprule
        \textbf{Model} &
        \textbf{Harness} &
        \textbf{Static Tool-Augmented Agent} &
        \textbf{ Internet-Enabled Open-World Agent} \\
        \midrule
        GPT-5.6 Luna
            & Codex    & \textbf{23\%} & \textbf{27\%} \\
            & OpenCode & 16\%          & 18\% \\
        \midrule
        GLM-5.3 Flash
            & ZCode    & \textbf{25\%} & 22\% \\
            & OpenCode & 24\%          & 22\% \\
        \bottomrule
    \end{tabular}
\end{table}

To isolate the effect of agent harness from that of the underlying model, we further compare GPT-5.6 Luna and GLM-5.3 Flash under both OpenCode and their default agent harness, while holding the model and analysis regime fixed.

The results in \autoref{tab:rq2_scaffold} show that harness can introduce additional performance variation even for the same base model. 
GPT-5.6 Luna performs consistently better with Codex than with OpenCode, whereas GLM-5.3 Flash is relatively insensitive to the choice of harness. 
This supports evaluating these systems as complete agent systems rather than attributing performance solely to the underlying model.

\end{document}